\pdfoutput=1
\documentclass[12pt,a4paper]{article}

\usepackage{ifthen} 
\newboolean{pdflatex}
\setboolean{pdflatex}{true} 

\newboolean{articletitles}
\setboolean{articletitles}{true} 

\newboolean{uprightparticles}
\setboolean{uprightparticles}{false} 

\usepackage{tikz-feynman}
\usepackage{subcaption}
\usepackage{booktabs}
\usepackage{multirow}

\def\paperauthors{LHCb collaboration} 
\def\paperasciititle{First observation and branching fraction measurement of the Lb->Dsp decay} 
\def\papertitle{First observation and branching fraction measurement of the $\LbDsP$ decay} 
\def\paperkeywords{{High Energy Physics}, {LHCb}} 
\def\papercopyright{\the\year\ CERN for the benefit of the LHCb collaboration} 
\def\paperlicence{CC BY 4.0 licence}
\def\paperlicenceurl{https://creativecommons.org/licenses/by/4.0/}

\usepackage[top=1in, bottom=1.25in, left=1in, right=1in]{geometry}

\usepackage{microtype}
\usepackage{lineno}  
\usepackage{xspace} 
\usepackage{caption} 

\usepackage{graphicx}  
\usepackage{color}
\usepackage{colortbl}
\graphicspath{{./figs/}} 

\usepackage{amsmath} 
\usepackage{amssymb}
\usepackage{amsfonts}
\usepackage{upgreek} 

\newcommand*\patchAmsMathEnvironmentForLineno[1]{%
\expandafter\let\csname old#1\expandafter\endcsname\csname #1\endcsname
\expandafter\let\csname oldend#1\expandafter\endcsname\csname
end#1\endcsname
 \renewenvironment{#1}%
   {\linenomath\csname old#1\endcsname}%
   {\csname oldend#1\endcsname\endlinenomath}%
}
\newcommand*\patchBothAmsMathEnvironmentsForLineno[1]{%
  \patchAmsMathEnvironmentForLineno{#1}%
  \patchAmsMathEnvironmentForLineno{#1*}%
}
\AtBeginDocument{%
\patchBothAmsMathEnvironmentsForLineno{equation}%
\patchBothAmsMathEnvironmentsForLineno{align}%
\patchBothAmsMathEnvironmentsForLineno{flalign}%
\patchBothAmsMathEnvironmentsForLineno{alignat}%
\patchBothAmsMathEnvironmentsForLineno{gather}%
\patchBothAmsMathEnvironmentsForLineno{multline}%
\patchBothAmsMathEnvironmentsForLineno{eqnarray}%
}

\usepackage{hyperxmp}

\usepackage[pdftex,
            pdfauthor={\paperauthors},
            pdftitle={\paperasciititle},
            pdfkeywords={\paperkeywords},
            pdfcopyright={Copyright (C) \papercopyright},
            pdflicenseurl={\paperlicenceurl}]{hyperref}

\usepackage[colorinlistoftodos,textsize=scriptsize]{todonotes}

\usepackage[bottom,flushmargin,hang,multiple]{footmisc}

\usepackage[all]{hypcap} 

\usepackage{xspace} 
\usepackage{upgreek}

\def\lhcb   {\mbox{LHCb}\xspace}

\def\MagUp {\mbox{\em Mag\kern -0.05em Up}\xspace}

\ifthenelse{\boolean{uprightparticles}}%
{

 \def\Ppi         {\ensuremath{\uppi}\xspace}                 
                  
 \def\Prho        {\ensuremath{\uprho}\xspace}

 \def\PDelta      {\ensuremath{\Delta}\xspace}                 
 \def\PXi         {\ensuremath{\Xi}\xspace}                 
 \def\PLambda     {\ensuremath{\Lambda}\xspace}                 
 \def\PSigma      {\ensuremath{\Sigma}\xspace}                 
 \def\POmega      {\ensuremath{\Omega}\xspace}                 
 \def\PUpsilon    {\ensuremath{\Upsilon}\xspace}
 \let\oldPi\Pi
 \def\PPi         {\ensuremath{\oldPi}\xspace}

 \def\PB      {\ensuremath{\mathrm{B}}\xspace}                 
                  
 \def\PD      {\ensuremath{\mathrm{D}}\xspace}

 \def\PK      {\ensuremath{\mathrm{K}}\xspace}

 \def\Pb      {\ensuremath{\mathrm{b}}\xspace}                 
 \def\Pc      {\ensuremath{\mathrm{c}}\xspace}

 \def\Pi      {\ensuremath{\mathrm{i}}\xspace}

 \def\Pp      {\ensuremath{\mathrm{p}}\xspace}

 \def\Ps      {\ensuremath{\mathrm{s}}\xspace}                 
                  
 \def\Pu      {\ensuremath{\mathrm{u}}\xspace}

 \def\thebaroffset{0.0em}
}
{

 \def\Ppi         {\ensuremath{\pi}\xspace}                 
                  
 \def\Prho        {\ensuremath{\rho}\xspace}

 \mathchardef\PDelta="7101
 \mathchardef\PXi="7104
 \mathchardef\PLambda="7103
 \mathchardef\PSigma="7106
 \mathchardef\POmega="710A
 \mathchardef\PUpsilon="7107
 \mathchardef\PPi="7105
                  
 \def\PB      {\ensuremath{B}\xspace}                 
                  
 \def\PD      {\ensuremath{D}\xspace}

 \def\PK      {\ensuremath{K}\xspace}

 \def\Pb      {\ensuremath{b}\xspace}                 
 \def\Pc      {\ensuremath{c}\xspace}

 \def\Pi      {\ensuremath{i}\xspace}

 \def\Pp      {\ensuremath{p}\xspace}

 \def\Ps      {\ensuremath{s}\xspace}                 
                  
 \def\Pu      {\ensuremath{u}\xspace}

 \def\thebaroffset{0.18em}
}
\newcommand{\offsetoverline}[2][\thebaroffset]{\kern #1\overline{\kern -#1 #2}}%

\makeatletter
\ifcase \@ptsize \relax
  \newcommand{\miniscule}{\@setfontsize\miniscule{4}{5}}
\or
  \newcommand{\miniscule}{\@setfontsize\miniscule{5}{6}}
\or
  \newcommand{\miniscule}{\@setfontsize\miniscule{5}{6}}
\fi
\makeatother

\DeclareRobustCommand{\optbar}[1]{\shortstack{{\miniscule (\rule[.5ex]{1.25em}{.18mm})}
  \\ [-.7ex] $#1$}}

\def\uquark    {{\ensuremath{\Pu}}\xspace}

\def\squark    {{\ensuremath{\Ps}}\xspace}

\def\cquark    {{\ensuremath{\Pc}}\xspace}

\def\bquark    {{\ensuremath{\Pb}}\xspace}

\def\pion   {{\ensuremath{\Ppi}}\xspace}
\def\piz    {{\ensuremath{\pion^0}}\xspace}
\def\pip    {{\ensuremath{\pion^+}}\xspace}
\def\pim    {{\ensuremath{\pion^-}}\xspace}

\def\rhomeson {{\ensuremath{\Prho}}\xspace}

\def\rhop     {{\ensuremath{\rhomeson^+}}\xspace}

\def\kaon    {{\ensuremath{\PK}}\xspace}

\def\KorKbar {\kern \thebaroffset\optbar{\kern -\thebaroffset \PK}{}\xspace}

\def\Kp      {{\ensuremath{\kaon^+}}\xspace}
\def\Km      {{\ensuremath{\kaon^-}}\xspace}

\def\Kstarp  {{\ensuremath{\kaon^{*+}}}\xspace}

\def\D       {{\ensuremath{\PD}}\xspace}

\def\DorDbar {\kern \thebaroffset\optbar{\kern -\thebaroffset \PD}\xspace}
\def\Dz      {{\ensuremath{\D^0}}\xspace}

\def\Dp      {{\ensuremath{\D^+}}\xspace}
\def\Dm      {{\ensuremath{\D^-}}\xspace}

\def\DpDm    {\ensuremath{\Dp {\kern -0.16em \Dm}}\xspace}

\def\Dstarp  {{\ensuremath{\D^{*+}}}\xspace}

\def\Dsp     {{\ensuremath{\D^+_\squark}}\xspace}
\def\Dsm     {{\ensuremath{\D^-_\squark}}\xspace}

\def\Dsmp    {{\ensuremath{\D^{\mp}_\squark}}\xspace}

\def\Dssm    {{\ensuremath{\D^{*-}_\squark}}\xspace}

\def\B       {{\ensuremath{\PB}}\xspace}

\def\BorBbar {\kern \thebaroffset\optbar{\kern -\thebaroffset \PB}\xspace}
\def\Bz      {{\ensuremath{\B^0}}\xspace}

\def\Bd      {{\ensuremath{\B^0}}\xspace}

\def\BdorBdbar {\kern \thebaroffset\optbar{\kern -\thebaroffset \Bd}\xspace}

\def\Bs      {{\ensuremath{\B^0_\squark}}\xspace}

\def\BsorBsbar {\kern \thebaroffset\optbar{\kern -\thebaroffset \Bs}\xspace}

\def\Y#1S{\ensuremath{\PUpsilon{(#1S)}}\xspace}

\def\proton      {{\ensuremath{\Pp}}\xspace}
\def\antiproton  {{\ensuremath{\overline \proton}}\xspace}

\def\Deltares    {{\ensuremath{\PDelta}}\xspace}

\def\Lz          {{\ensuremath{\PLambda}}\xspace}
\def\Lbar        {{\ensuremath{\offsetoverline{\PLambda}}}\xspace}
\def\LorLbar     {\kern \thebaroffset\optbar{\kern -\thebaroffset \PLambda}\xspace}

\def\Lc          {{\ensuremath{\Lz^+_\cquark}}\xspace}
\def\Lcbar       {{\ensuremath{\Lbar{}^-_\cquark}}\xspace}

\def\Lb           {{\ensuremath{\Lz^0_\bquark}}\xspace}

\def\BF         {{\ensuremath{\mathcal{B}}}\xspace}

\newcommand{\decay}[2]{\ensuremath{#1\!\to #2}\xspace} 

\def\to                 {\ensuremath{\rightarrow}\xspace}

\def\Vcs  {{\ensuremath{V_{\cquark\squark}^{\phantom{\ast}}}}\xspace}

\def\Vub  {{\ensuremath{V_{\uquark\bquark}^{\phantom{\ast}}}}\xspace}

\def\AT#1     {\ensuremath{A_{\mathrm{T}}^{#1}}\xspace}           

\def\C#1      {\ensuremath{\mathcal{C}_{#1}}\xspace}                       
\def\Cp#1     {\ensuremath{\mathcal{C}_{#1}^{'}}\xspace}                    
\def\Ceff#1   {\ensuremath{\mathcal{C}_{#1}^{\mathrm{(eff)}}}\xspace}        
\def\Cpeff#1  {\ensuremath{\mathcal{C}_{#1}^{'\mathrm{(eff)}}}\xspace}       
\def\Ope#1    {\ensuremath{\mathcal{O}_{#1}}\xspace}                       
\def\Opep#1   {\ensuremath{\mathcal{O}_{#1}^{'}}\xspace}                    

\newcommand{\nospaceunit}[1]{\ensuremath{\text{#1}}}       
\newcommand{\aunit}[1]{\ensuremath{\text{\,#1}}}       

\newcommand{\tev}{\aunit{Te\kern -0.1em V}\xspace}
\newcommand{\gev}{\aunit{Ge\kern -0.1em V}\xspace}
\newcommand{\mev}{\aunit{Me\kern -0.1em V}\xspace}
\newcommand{\kev}{\aunit{ke\kern -0.1em V}\xspace}
\newcommand{\ev}{\aunit{e\kern -0.1em V}\xspace}
 
\newcommand{\mevc}{\ensuremath{\aunit{Me\kern -0.1em V\!/}c}\xspace}
\newcommand{\gevc}{\ensuremath{\aunit{Ge\kern -0.1em V\!/}c}\xspace}
\newcommand{\mevcc}{\ensuremath{\aunit{Me\kern -0.1em V\!/}c^2}\xspace}
\newcommand{\gevcc}{\ensuremath{\aunit{Ge\kern -0.1em V\!/}c^2}\xspace}

\def\mum  {\ensuremath{\,\upmu\nospaceunit{m}}\xspace}

\def\fb   {\ensuremath{\aunit{fb}}\xspace}
\def\invfb   {\ensuremath{\fb^{-1}}\xspace}

\newcommand{\chisq}{\ensuremath{\chi^2}\xspace}

\newcommand{\chisqip}{\ensuremath{\chi^2_{\text{IP}}}\xspace}

\def\gsim{{~\raise.15em\hbox{$>$}\kern-.85em
          \lower.35em\hbox{$\sim$}~}\xspace}
\def\lsim{{~\raise.15em\hbox{$<$}\kern-.85em
          \lower.35em\hbox{$\sim$}~}\xspace}

\def\sqs   {\ensuremath{\protect\sqrt{s}}\xspace}

\def\pt         {\ensuremath{p_{\mathrm{T}}}\xspace}

\def\ptot       {\ensuremath{p}\xspace}

\def\evtgen     {\mbox{\textsc{EvtGen}}\xspace}

\def\geant      {\mbox{\textsc{Geant4}}\xspace}

\def\photos     {\mbox{\textsc{Photos}}\xspace}

\def\pythia     {\mbox{\textsc{Pythia}}\xspace}

\def\tell1  {TELL1\xspace}
\def\ukl1   {UKL1\xspace}

\newcommand{\ie}{\mbox{\itshape i.e.}\xspace}

\newcommand{\lhcborcid}[1]{\href{https://orcid.org/#1}{\hspace*{0.1em}\raisebox{-0.45ex}{\includegraphics[width=1em]{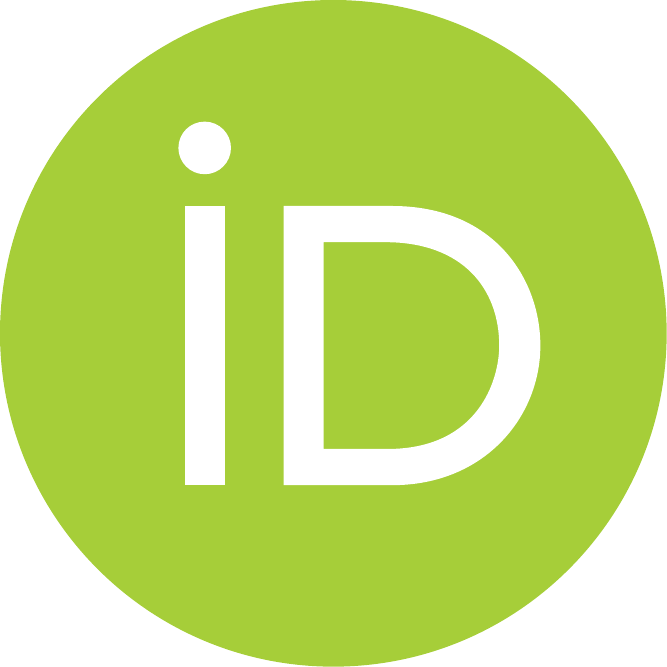}}}}

\usepackage{cite} 
\usepackage{mciteplus}

\def\Deltap      {{\ensuremath{\Deltares^{+}}}\xspace}

\newcommand{\BdDK}     {\texorpdfstring{\decay{\Bz}{\Dm \Kp}}{}}
\newcommand{\BdDPi}    {\texorpdfstring{\decay{\Bz}{\Dm \pip}}{}}

\newcommand{\BdDsK}    {\texorpdfstring{\decay{\Bz}{\Dsm \Kp}}{}}
\newcommand{\BdDsPi}   {\texorpdfstring{\decay{\Bz}{\Dsp \pim}}{}}

\newcommand{\BsDsK}    {\texorpdfstring{\decay{B^0_s}{D_s^\mp K^\pm}}{}}
\newcommand{\BsDsPi}   {\texorpdfstring{\decay{B^0_s}{D_s^- \pi^+}}{}}

\newcommand{\BsDsRho}  {\decay{\Bs}{\Dsm \rho^{+}}}

\newcommand{\BsDsstPi} {\decay{\Bs}{\Dssm \pip}}
\newcommand{\BdDsstPi} {\decay{\Bz}{\Dssm \pip}}
\newcommand{\BdDsstK} {\decay{\Bz}{\Dssm  \Kp}}
\newcommand{\BsDsstRho} {\decay{\Bs}{\Dssm \rho^{+}}}
\newcommand{\DKPiPi}   {\decay{\Dm}{\Kp\pim\pim}}
\newcommand{\DsKKPi}   {\decay{\Dsm}{\Km\Kp\pim}}

\newcommand{\BsDsPilike} {\decay{\Bs}{D_s^{(*)-}\{\pi^+,\rho^+\}}}
\newcommand{\BsDsKlike} {\decay{B_{(s)}^0}{D_s^{(*)\mp}K^{(*)\pm}}}

\newcommand{\BsDsKst}    {\texorpdfstring{\decay{B_s^0}{D_s^\mp K^{*\pm}}}{}}
\newcommand{\BsDsstK}    {\texorpdfstring{\decay{B_s^0}{D_s^{*\mp} K^\pm}}{}}
\newcommand{\BsDsstKst}    {\texorpdfstring{\decay{B_s^0}{D_s^{*\mp} K^{*\pm}}}{}}
\newcommand{\LbLcRho} {\texorpdfstring{\decay{\Lb}{\Lc \rho^- }}{}}

\newcommand{\DsPi}     {\texorpdfstring{D_s^- \pi^+}{}}

\newcommand{\LbDsP}    {\texorpdfstring{\decay{\Lb}{\Dsm p}}{}}

\newcommand{\LbDsstP}  {\decay{\Lb}{\Dssm p}}
\newcommand{\LbDsDelta}{\decay{\Lb}{\Dsm \Deltap}}
\newcommand{\LbLcK}    {\decay{\Lb}{\Lc \Km}}
\newcommand{\LbLcPi}   {\decay{\Lb}{\Lc \pim}}

\newcommand{\LcPKPi}   {\decay{\Lc}{\proton \Km \pip}}

\newcommand{\DsP}      {\texorpdfstring{\Dsm\Pp}{}}

\newcommand{\DPi}      {\texorpdfstring{D^- \pi^+}{}}

\newcommand{\LcPi}     {\texorpdfstring{\Lc \pim}{}}

\newcommand{\mLbLcPi} {\texorpdfstring{m(\LcPi)}{}}
\newcommand{\mLbDsP} {\texorpdfstring{m(\DsP)}{}}
\newcommand{\mKKPi} {\texorpdfstring{m(\Km\Kp\pim)}{}}

\newcommand{\aNF}{\ensuremath{a_{N\!F}}}

\newcommand{\BFLbDspPaper}{(12.6 \pm 0.5 \pm 0.3 \pm 1.2 )\times 10^{-6}}

\newcommand{\RatioBFPAPER}{(2.56 \pm 0.10 \pm 0.05 \pm 0.14 )\times 10^{-3}}

\newcommand{\YieldLbDsp}{831\pm32}

\newcommand{\YieldLbLcPiPaper}{(4.047\pm0.007)\times 10^5}

\newcommand{\EffLbDspPerc}{0.1819 \pm 0.0013}

\newcommand{\EffLbLcPiPerc}{0.1947 \pm 0.0012}

\newcommand{\EffRatioLbLcPiLbDspPaper}{1.070\phantom{0} \pm 0.010\phantom{0}}

\usepackage{longtable} 

\begin{document}

\renewcommand{\thefootnote}{\fnsymbol{footnote}}
\setcounter{footnote}{1}


\begin{titlepage}
\pagenumbering{roman}

\vspace*{-1.5cm}
\centerline{\large EUROPEAN ORGANIZATION FOR NUCLEAR RESEARCH (CERN)}
\vspace*{1.5cm}
\noindent
\begin{tabular*}{\linewidth}{lc@{\extracolsep{\fill}}r@{\extracolsep{0pt}}}
\ifthenelse{\boolean{pdflatex}}
{\vspace*{-1.5cm}\mbox{\!\!\!\includegraphics[width=.14\textwidth]{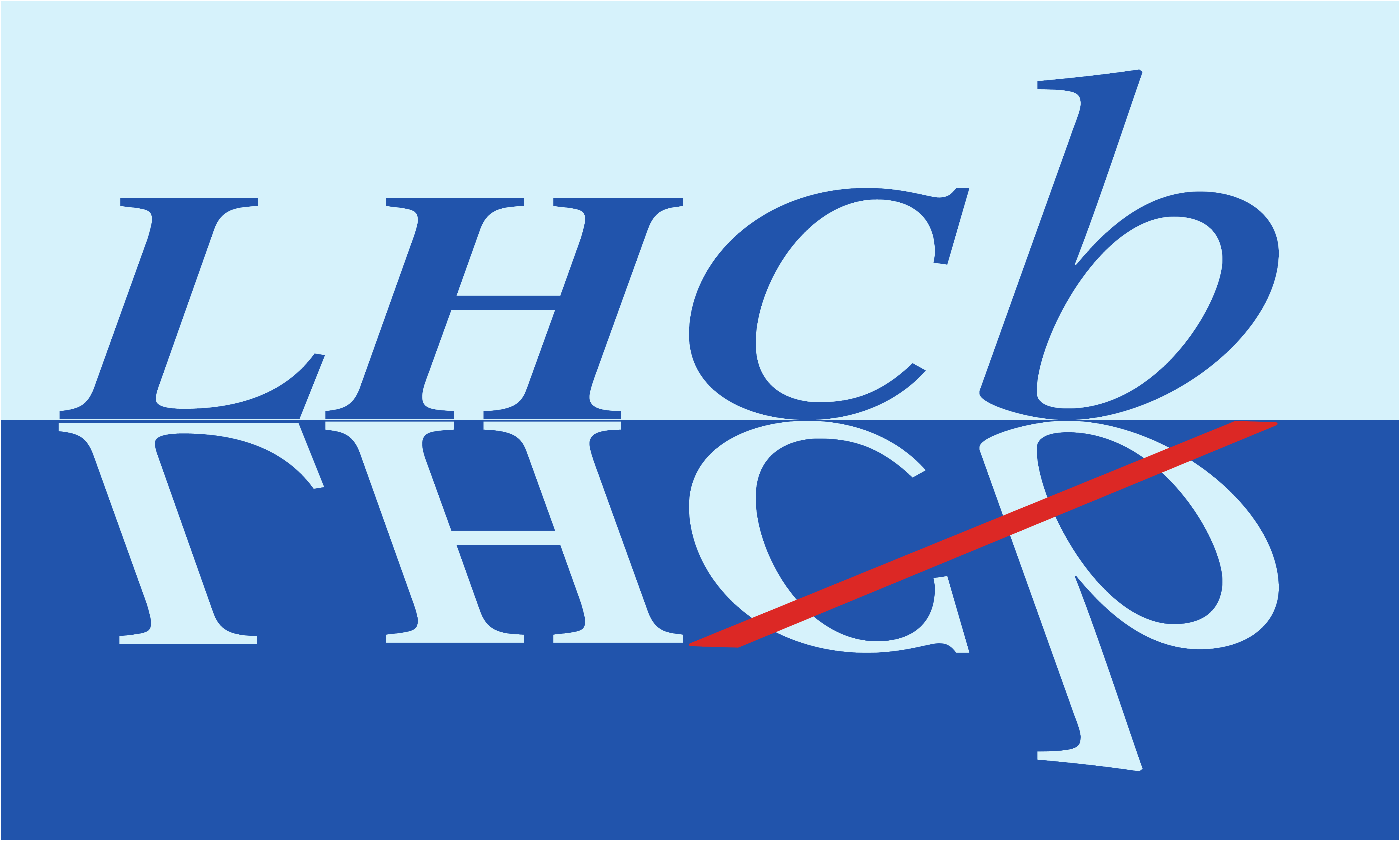}} & &}%
{\vspace*{-1.2cm}\mbox{\!\!\!\includegraphics[width=.12\textwidth]{figs/lhcb-logo.eps}} & &}%
\\
 & & CERN-EP-2022-272 \\  
 & & LHCb-PAPER-2022-038 \\  
 & & 17 July 2023 \\ 
 & & \\
\end{tabular*}

\vspace*{4.0cm}

{\normalfont\bfseries\boldmath\huge
\begin{center}
  \papertitle 
\end{center}
}

\vspace*{2.0cm}

\begin{center}
\paperauthors\footnote{Authors are listed at the end of this paper.}
\end{center}

\vspace{\fill}

\begin{abstract}
  \noindent
  The first observation of the \LbDsP decay is presented using proton-proton collision data collected by the \lhcb experiment at a centre-of-mass energy of ${\sqs=13\tev}$, corresponding to a total integrated luminosity of $6\invfb$. Using the $\LbLcPi$ decay as the normalisation mode, the branching fraction of the \mbox{$\LbDsP$} decay is measured to be ${\BF(\LbDsP)=\BFLbDspPaper}$, where the first uncertainty is statistical, the second systematic and the third due to uncertainties in the branching fractions of the $\LbLcPi$, $\DsKKPi$ and $\LcPKPi$ decays.
\end{abstract}

\vspace*{2.0cm}

\begin{center}
   Published in JHEP 07 (2023) 075 
\end{center}

\vspace{\fill}

{\footnotesize 
\centerline{\copyright~\papercopyright. \href{\paperlicenceurl}{\paperlicence}.}}
\vspace*{2mm}

\end{titlepage}


\newpage
\setcounter{page}{2}
\mbox{~}
%
%
%
%


\renewcommand{\thefootnote}{\arabic{footnote}}
\setcounter{footnote}{0}

\cleardoublepage


\pagestyle{plain} 
\setcounter{page}{1}
\pagenumbering{arabic}



\section{Introduction}
\label{sec:introduction}

The Standard Model (SM) of particle physics describes the elementary particles and their interactions.
In the SM, the Cabibbo-Kobayashi-Maskawa (CKM) mechanism describes how the weak interaction eigenstates are related to the mass eigenstates of the quarks and, therefore, also describes the interaction strengths among quarks via the weak interaction~\cite{Cabibbo:1963yz,Kobayashi:1973fv}.
The CKM-matrix element describing the $\bquark\to\uquark$ transition, $\Vub$, is the element with the smallest and most poorly determined magnitude.  
Better knowledge on $|\Vub|$ provides a critical input to check the consistency of the SM~\cite{CKMfitter2015, UTfit-UT}.

\begin{figure}[b]
    \centering
    \vspace{+0.1cm}
    \includegraphics[width=0.60\textwidth]{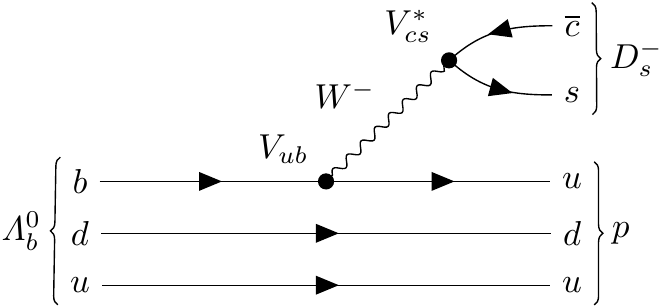}
    \caption{
    Tree diagram contributing to the \LbDsP decay.
    }
    \label{fig:feyn-lb2dsp}
\end{figure}

The $\LbDsP$ decay\footnote{Inclusion of charge-conjugated modes is implied unless explicitly stated.} is a weak hadronic decay that proceeds through a $\bquark \to \uquark$ transition.
A single leading-order diagram contributes to this process, shown in Fig.~\ref{fig:feyn-lb2dsp}.
Therefore, the $\LbDsP$ branching fraction is proportional to $|\Vub|^2$,
\begin{equation}
    \BF(\LbDsP) \propto |\Vub|^{2} |\Vcs|^{2} f_{D_s}^{2} |\aNF|^{2} |F_{\Lb \rightarrow \proton}(m_{D_s}^2)|^{2}\ ,
\label{eq:factorisation}
\end{equation}
where $|\Vcs|$ describes the $\cquark \to \squark$ quark transition,  $f_{D_s}$ is the $\Dsm$ decay constant, $F_{\Lb \rightarrow \proton}$ is the form factor describing the $\Lb$ to proton transition and $|\aNF|$ quantifies the nonfactorisable effects, \ie final-state strong interactions leading to a nonfactorisable contribution in the decay amplitude.
The $f_{D_s}$ constant has been calculated with high precision using lattice QCD~\cite{Bazavov:2017lyh, Carrasco:2014poa} and $|\Vcs|$ is known to be close to unity~\cite{PDG2022}.
Describing a heavy-quark decay as the product of the matrix elements of two currents is known as factorisation and leads to the product of the decay constant and the form factor.
Factorisation is expected to be a good approximation if the emitted meson is light relative to the other hadron, formed from the spectator quarks~\cite{Beneke:2000ry}.
Therefore, factorisation may not hold in the $\LbDsP$ decay, where the $\Dsm$ meson is heavier than the proton.

The calculations of the matrix elements simplify in semileptonic decays, where they are the product of the leptonic and the hadronic currents.
Therefore, semileptonic decays are used to determine $|\Vub|$, where the hadronic matrix elements can be calculated using nonperturbative methods.
By contrast, the knowledge of the matrix elements in hadronic decays is poorly known, as they are plagued by nonfactorisable effects in the final state.
Recently, the branching fraction of the $\BdDsPi$ decay has been measured~\cite{LHCb-PAPER-2020-021}.
The $\BdDsPi$ decay proceeds with the same tree-level transition as the $\LbDsP$ decay, leading to the same expression as Eq.~\ref{eq:factorisation}, except for the form factor and nonfactorisable effects.
For the $\BdDsPi$ decay, the form factor is known with about $10\%$ precision~\cite{formfactor,Ball:2006jz}, which allows for the determination of the product of $|\Vub|$ and $|\aNF|$.
The $\LbDsP$ decay can, in principle, provide another measure of the breaking of the factorisation hypothesis.
However, the form factor for this decay is not accurately known yet~\cite{detmold2015lambda}.
Therefore, this measurement provides an incentive to address these calculations.

This paper presents the first observation and branching fraction measurement of the \mbox{$\LbDsP$} decay using proton-proton ($pp$) collision data collected with the LHCb detector at a centre-of-mass energy of $13\tev$ and corresponding to an integrated luminosity of $6\invfb$.
Data taken in Run 2 of the Large Hadron Collider (LHC) between 2015 and 2018 is used.
The $\LbLcPi$ decay is used as a normalisation channel because it is topologically similar to the signal decay and has a relatively high branching fraction.
Candidates of $\LbDsP$ ($\LbLcPi$) decays are reconstructed using the final-state particles of the $\DsKKPi$ ($\LcPKPi$) decay.
The branching fraction of $\LbDsP$ is determined using
\begin{equation}
    \BF(\LbDsP) = \BF(\LbLcPi) \dfrac{N_{\LbDsP}}{N_{\LbLcPi}} \dfrac{\epsilon_{\LbLcPi}}{\epsilon_{\LbDsP}} \dfrac{\BF(\LcPKPi)}{\BF(\DsKKPi)}\ ,
    \label{eq:BF}
\end{equation}
where $N_{X}$ is the measured yield of decay $X$ and $\epsilon_{X}$ is the efficiency of the candidate reconstruction and selection.
The candidate yields are determined using unbinned extended maximum-likelihood fits of the invariant mass of the $\Lb$ hadron.
The corresponding efficiencies are determined using simulated candidates and calibration data samples.

\section{Detector and simulation}
\label{sec:detector}

The \lhcb detector~\cite{LHCb-DP-2008-001,LHCb-DP-2014-002} is a single-arm forward
spectrometer covering the \mbox{pseudorapidity} range $2<\eta <5$,
designed for the study of particles containing \bquark or \cquark
quarks. The detector includes a high-precision tracking system
consisting of a silicon-strip vertex detector surrounding the $pp$
interaction region, a large-area silicon-strip detector located
upstream of a dipole magnet with a bending power of about
$4{\mathrm{\,Tm}}$, and three stations of silicon-strip detectors and straw
drift tubes placed downstream of the magnet.
The tracking system provides a measurement of the momentum, \ptot, of charged particles with
a relative uncertainty that varies from 0.5\% at low momentum to 1.0\% at 200\gevc.
The minimum distance of a track to a primary $pp$ collision vertex (PV), the impact parameter (IP), 
is measured with a resolution of $(15+29/\pt)\mum$,
where \pt is the component of the momentum transverse to the beam, in\,\gevc.
Different types of charged hadrons are distinguished using information
from two ring-imaging Cherenkov detectors. 
Photons, electrons and hadrons are identified by a calorimeter system consisting of
scintillating-pad and preshower detectors, an electromagnetic
and a hadronic calorimeter. Muons are identified by a
system composed of alternating layers of iron and multiwire
proportional chambers.
The online event selection is performed by a trigger, 
which consists of a hardware stage, based on information from the calorimeter and muon
systems, followed by a software stage, which applies a full event
reconstruction.
At the hardware trigger stage, events are required to have a muon with high \pt or a
hadron, photon or electron with high transverse energy in the calorimeters. For hadrons,
the transverse energy threshold is 3.5\gev.
Events selected by the hadron trigger or by any other trigger independently of the signal candidate are used in this analysis.
The software trigger requires a two-, three- or four-track
secondary vertex with a significant displacement from any primary
$pp$ interaction vertex. At least one charged particle
must have a transverse momentum $\pt > 1.6\gevc$ and be
inconsistent with originating from a PV.
A multivariate algorithm~\cite{BBDT,LHCb-PROC-2015-018} is used for
the identification of secondary vertices consistent with the decay
of a \bquark hadron.

Simulation is required to calculate reconstruction and selection efficiencies, and to determine shapes of invariant-mass distributions.
In the simulation, $pp$ collisions are generated using
\pythia~\cite{Sjostrand:2007gs} 
with a specific \lhcb configuration~\cite{LHCb-PROC-2010-056}.
Decays of unstable particles
are described by \evtgen~\cite{Lange:2001uf}, in which final-state
radiation is generated using \photos~\cite{davidson2015photos}.
The interaction of the generated particles with the detector, and its response,
are implemented using the \geant
toolkit~\cite{Allison:2006ve, *Agostinelli:2002hh} as described in
Ref.~\cite{LHCb-PROC-2011-006}. 

\section{Selection of candidates}
\label{sec:selection}

The $\LbDsP$ (\LbLcPi) decay is reconstructed by selecting \mbox{$\DsKKPi$} (\mbox{$\LcPKPi$}) candidates and combining them with a proton (charged pion), which is referred to as the companion particle.
A kinematic fit~\cite{Hulsbergen:2005pu} is used to improve the $\Lb$ mass resolution by constraining the $\Dsm$ ($\Lcbar$) mass to its known value~\cite{PDG2022}.
The $\DsP$ (\LcPi) invariant mass, $\mLbDsP$ ($\mLbLcPi$), is required to be within the range $5200\text{--}6200$ $(5400\text{--}6200)\mevcc$ and the unconstrained $\Km\Kp\pim$ ($\antiproton\Kp\pim$) invariant mass must be within $1948\text{--}1988$ $(2266\text{--}2306)\mevcc$. 

Candidates that have been selected by the trigger requirements, as described in the previous section, are subject to further offline selection to reduce the background contributions.
First, a preselection reconstructs $b$-hadron and $c$-hadron candidates with good-quality vertices by reconstructing four well-reconstructed tracks with high transverse and total momentum and inconsistent with originating from any PV.
A gradient-boosted decision tree (BDTG) algorithm~\cite{Breiman, Roe:2004na} is used to reduce the background contributions due to random combinations of final-state particles. This BDTG classifier is trained on $\BsDsPi$ candidates taken in 2011 and 2012 (Run~1) and is described in Ref.~\cite{LHCb-PAPER-2017-047}.
The BDTG is suitable for decays topologically similar to $\BsDsPi$, as it does not use particle identification variables.
The BDTG classifier combines a number of track-related variables, including the transverse momentum of the companion particle, the $\bquark$-hadron and $\cquark$-hadron candidate's radial flight distance and the companion and $\bquark$ hadron's minimum $\chisqip$, where $\chisqip$ is defined as the difference in the vertex-fit $\chisq$ of the PV reconstructed with and without the candidate.
A more detailed description of the BDTG classifier is given in Ref.~\cite{Eitschberger:2018ofp}.
It has been verified using $\LbDsP$ simulation that this BDTG performs well in separating signal from combinatorial background.

To separate $\LbDsP$ from backgrounds with a misidentified final-state particle, a requirement on the PID of the companion proton is applied. 
This requirement is based on multivariate techniques to combine the PID and tracking information of all LHCb subsystems to get a single discriminating variable~\cite{LHCb-PUB-2016-021}.
The optimal requirements for the BDTG classifier and the companion identification variables are simultaneously determined by maximising the figure of merit $\epsilon_{\textrm{sig}} / (5 + \sqrt{N_{\textrm{bg}}})$~\cite{Punzi:2003bu} expected for the signal region ($\mLbDsP \in [5580, 5660] \mevcc$).
Here, $\epsilon_{\textrm{sig}}$ is the signal efficiency of the corresponding BDTG and PID requirements in addition to the other selection requirements, and $N_{\textrm{bg}}$ is the expected number of background events in the signal region for the given set of selection requirements.
The optimal point of this figure of merit corresponds to a tight PID requirement on the companion proton with a signal efficiency of $76\%$, calculated after applying the other selection requirements.
For the $\LbLcPi$ sample, similar to the signal sample, a PID requirement is imposed on the companion pion to remove contributions from $\LbLcK$ decays.

To further reduce contributions of backgrounds due to misidentification of the final-state particles, PID requirements are imposed on the decay products of the $\cquark$-hadron.
Furthermore, vetoes are applied to the $\Dsmp h^{\pm}$ samples to reduce the contribution from $\DKPiPi$ and $\LcPKPi$ decays misidentified as $\DsKKPi$ candidates.
In these vetoes, the $\Dsm$ candidates are reconstructed as $\DKPiPi$ or $\LcPKPi$ decays, and a tight PID requirement is imposed on those near the known $\Dm$ or $\Lcbar$ mass.
The PID requirements together have a signal efficiency of $56\%$ for the $\LbDsP$ sample and $71\%$ for the $\LbLcPi$ sample, calculated after applying the other selection requirements.
Finally, contributions from $\bquark$-hadron decays not including a charm hadron are suppressed by a requirement on the $\cquark$-hadron's flight-distance significance, defined as the distance between the \bquark- and \cquark-hadron decay vertices divided by the uncertainty on this measurement. This requirement is applied to all samples.

The efficiency of the candidate selection is calculated using simulated $\LbDsP$ and $\LbLcPi$ decays, except for the PID and hardware-trigger efficiencies, which are determined using calibration data samples.
The efficiency of the PID requirements is calculated using samples of $\Dstarp\to\Dz\pip$ and $\Lc\to\proton\Km\pip$ calibration data. 
A more complete description of this method can be found in Ref.~\cite{LHCb-DP-2018-001}.
The determination of the hardware trigger efficiency uses the $\Dstarp\to\Dz\pip$ and $\Lz^0\to\proton\pim$ calibration samples and simulated signal. 
The efficiencies of hadrons to be triggered depend on and therefore are calibrated based on the type and charge of the particle, as well as the magnitude and the calorimeter region, demarcated by the cell sizes, of the corresponding energy deposit in the calorimeter.
Table~\ref{tab:eff-total} shows the ratio of the selection efficiencies between the signal and normalisation channels.
The most sizeable differences in efficiency are due to the different PID requirements for the signal and normalisation channel and due to the flight-distance significance and momentum requirements.

\begin{table}
\small
\centering
\begin{footnotesize} 
\caption{A breakdown of the relative efficiency ratios of the \LbLcPi and \LbDsP decays, calculated after applying the preceding requirements. The uncertainty on the efficiencies is due to the size of the simulation and calibration samples.}
\label{tab:eff-total}
\begin{tabular}{lc}
  \toprule
	Requirement      & Ratio $\epsilon(\LbLcPi)$/$\epsilon(\LbDsP)$  \\
	\midrule 
	LHCb acceptance           & $0.9625 \pm 0.0016$ \\
	Software trigger and preselection &  $1.1370 \pm 0.0026$ \\
	Kinematic and geometric selection &  $0.7580 \pm 0.0018$ \\
	Particle identification selection &  $1.278\phantom{0} \pm 0.005\phantom{0}$  \\
 	Hardware trigger          &  $0.995\phantom{0} \pm 0.006\phantom{0}$  \\
	\midrule
	Total                     &  $\EffRatioLbLcPiLbDspPaper$ \\
  \bottomrule

\end{tabular}
\end{footnotesize} 
\end{table} 

Two control channels, $\BsDsPi$ and $\BsDsK$, are used to estimate the contributions of misidentified $\BsDsPilike$ and $\BsDsKlike$ decays in the $\LbDsP$ sample.
The control modes are subjected to the same candidate selection as $\LbDsP$ candidates, except for the particle identification (PID) requirement on the companion particle.

\section{Invariant-mass fits}
\label{sec:mass-fits}

The yields of the signal $\LbDsP$ and normalisation $\LbLcPi$ channels are determined using unbinned maximum-likelihood fits to the $\Dsm\proton$ and $\LcPi$ invariant-mass distributions, respectively.
The candidate samples from different years of data-taking and magnet polarities are combined in the fits.

The signal components in the invariant-mass fits are parameterised using the sum of a double-sided Hypatia function~\cite{Santos:2013gra} and a Johnson $S_U$ function~\cite{johnson1949systems}. 
Their corresponding parameters are obtained from fits to samples of simulated candidates.
In the fit to $\LbLcPi$ candidates in data, the common mean of the functions and the widths are left unconstrained to account for any under- or overestimation of resolution effects in the simulation.
The widths from the fit to $\LbLcPi$ candidates, scaled with the expected width differences from simulation, are used to fix the widths in the fit to $\LbDsP$ candidates, where only the mean is left unconstrained.

The residual combinatorial background contribution is modelled using analytic functions.
In the $\mLbDsP$ fit, this contribution is parametrised as \mbox{$\mathcal{C}(m|m_0, A, C) = \left(1 - \exp(-(m-m_0)/C)\right)\times \left( m/m_0\right)^A,$} which describes the turn-on point in the $\Dsm\proton$ invariant-mass due to $\Dsm\pip$ invariant-mass requirements in the preselection.
The parameters of this function are determined by fitting the $\mLbDsP$ distribution using the upper sideband region of $\mKKPi$ and are subsequently fixed in the fit to data.
In the $\mLbLcPi$ fit, the combinatorial background is described using an exponential distribution, validated on the upper $\mLbLcPi$ sideband.
The exponential parameter is left unconstrained in the fit to account for kinematic differences between the $\mLbLcPi$ sideband and the full range.

Decays where one or more of the final-state particles are missed by the reconstruction are referred to as partially reconstructed backgrounds.
In the fits to the signal and normalisation channels, these backgrounds are parametrised by parabolic functions convolved with a double Gaussian distribution to account for detector effects.
These functions describe decays where a neutral pion or photon is not reconstructed~\cite{LHCb-PAPER-2017-021}.
The fit to $\LbDsP$ candidates considers partially reconstructed background components from $\Lb\rightarrow \Dssm (\rightarrow \Dsm \gamma/\piz)\proton$ decays.
As the $\Dssm$ meson decays to either $\Dsm\gamma$ or to $\Dsm\piz$, this contribution is described by a combination of two of the above-mentioned parabolic functions.
The corresponding parameters are obtained from simulation and fixed in the fit to data.
The yield of $\LbDsstP$ decays is left unconstrained in the fit to $\LbDsP$ candidates, as there is no estimate of the corresponding branching fraction.
The addition of an extra partially reconstructed component describing $\LbDsDelta$ decays has been studied and is found to have a negligible impact on the signal yield.
The $\LbLcPi$ sample contains partially reconstructed backgrounds from $\LbLcRho (\rightarrow \pim \piz)$ and $\Lb \rightarrow \Sigma_c^+(\rightarrow \Lc \piz) \pim $ decays.
These are both parametrised using a parabolic function to describe the missing neutral pion.
The parameters of this function are obtained from simulation and fixed in the $\mLbLcPi$ fit, while the yields are left free.

The background contributions due to the misidentification of the companion particle in the $\mLbDsP$ fit consist of the decays $\BsDsPi$, $\BsDsK$, $\BdDsK$ and the corresponding backgrounds with missing photons or neutral pions in the final state, originating from $\rhop\to\pip\piz$, $\Kstarp\to\Kp\piz$ or $\Dssm\to\Dsm\{\gamma,\piz\}$ decays.
The shapes of the misidentified backgrounds in the $\mLbDsP$ fit are determined from simulation using a nonparametric kernel estimation method~\cite{Cranmer:2000du}.
The exceptions are the $\BsDsstPi$ and $\BdDsstPi$ components, described by Crystal Ball functions~\cite{Skwarnicki:1986xj}.
The shape of the $\BsDsstPi$ contribution is obtained from simulation.
The same parametrisation shifted by the known $\Bd \text{--} \Bs$ mass difference is used to model $\BdDsstPi$  decays.
All misidentified backgrounds have fixed shapes in the fit to the $\Dsm\proton$ invariant mass.
Fits to the $\Bs$ invariant mass in the $\BsDsPi$ and $\BsDsK$ control samples (see Appendix~\ref{app:control-channels}) provide an estimate of the contributions of the misidentified background components in the $\LbDsP$ sample.
These estimates are computed by correcting the observed yields of $\BsDsPilike$ and $\BsDsKlike$ decays for the different PID requirements between the control and signal samples.
Subsequently, they are constrained in the $\mLbDsP$ fit.

The sample of $\LbLcPi$ candidates is contaminated by the $\LbLcK$, $\BsDsPi$ and $\BdDPi$ backgrounds due to the misidentification of one of the final-state particles.
The shapes of these backgrounds are determined from simulation using a nonparametric kernel estimation method and fixed in the fit to data.
The size of the $\LbLcK$ contribution is constrained to the expected yield determined using knowledge of its branching fraction~\cite{LHCB-PAPER-2013-056} and efficiencies obtained from simulation.
A data-driven method is used to determine the $\BsDsPi$ and $\BdDPi$ yields in the $\mLbLcPi$ fit.
The $\LcPi$ data is reconstructed as $\DsPi$ and $\DPi$, fitted, and the resulting yields are corrected for the difference in PID and invariant-mass requirements. 
Ultimately, the number of expected $\BsDsPi$ and $\BdDPi$ is small and is therefore fixed in the fit to $\LbLcPi$ candidates.

\begin{figure}[htp]
   \begin{center}
        \includegraphics[width=0.66\textwidth]{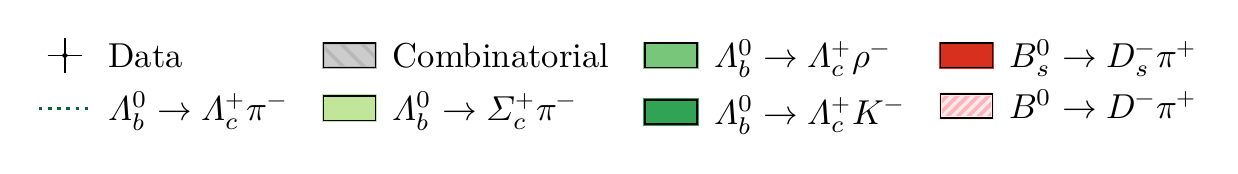}\\
        \includegraphics[width=0.49\textwidth]{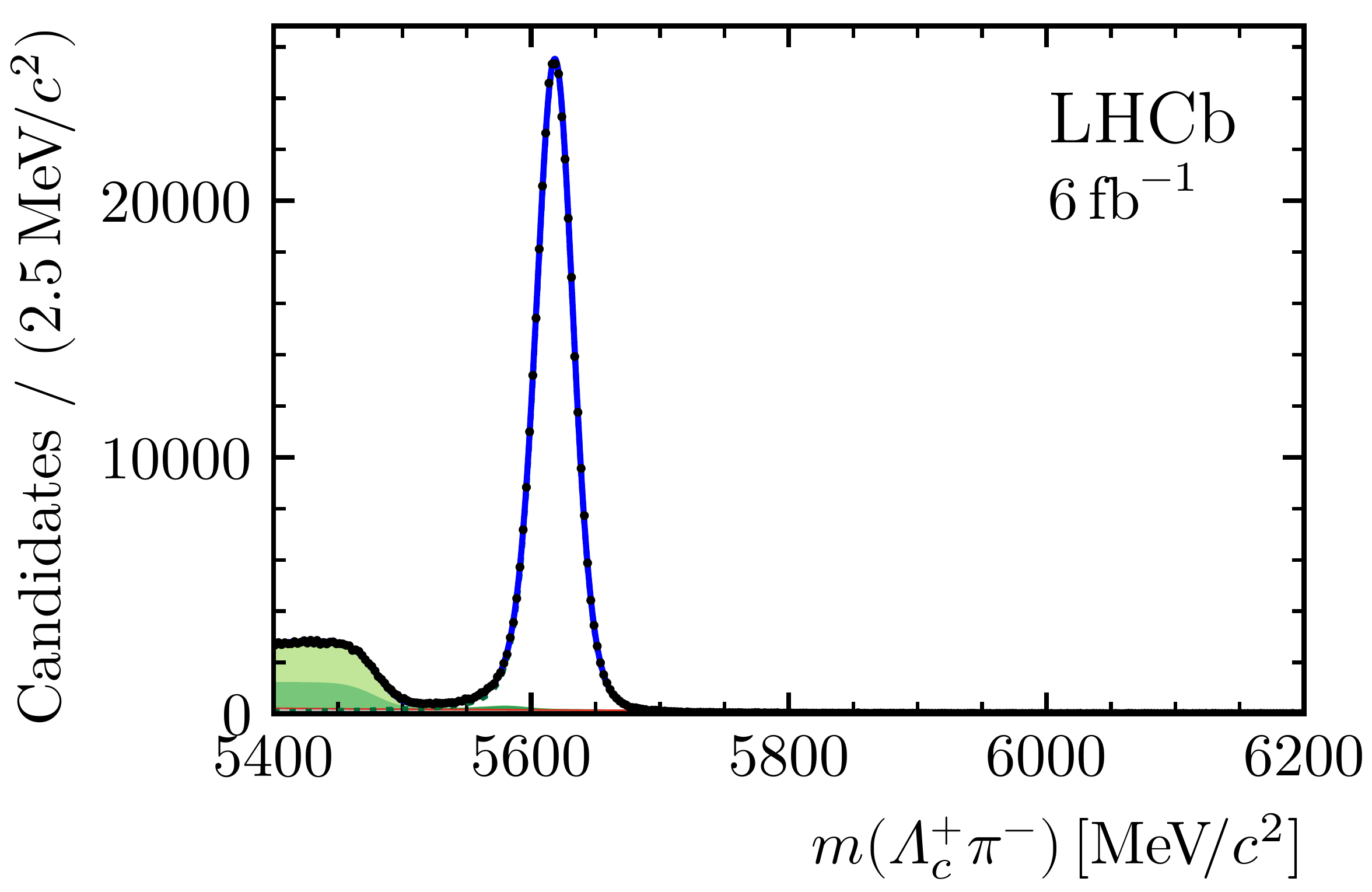}
        \includegraphics[width=0.49\textwidth]{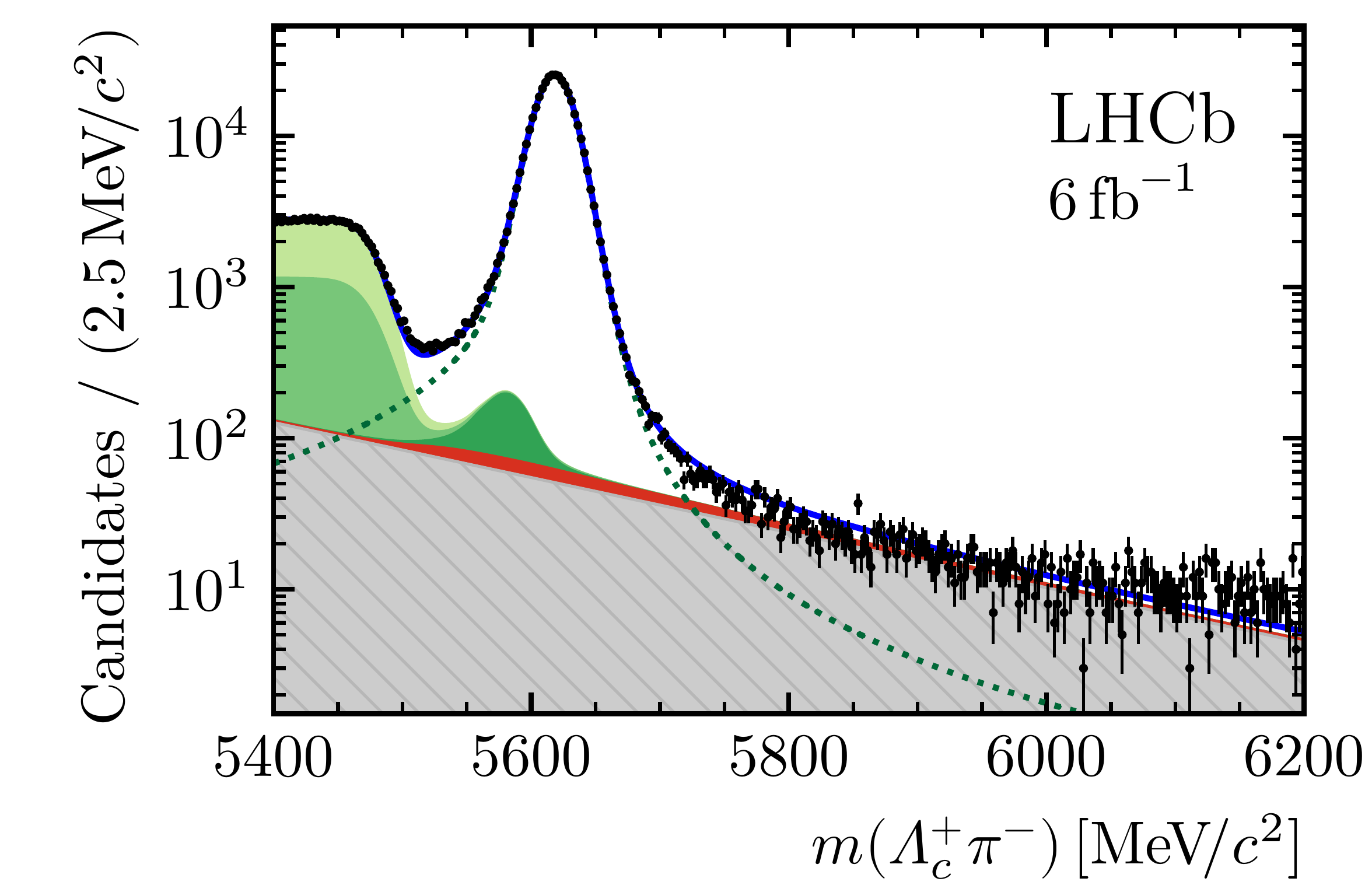}
    \end{center}
	\caption{Invariant-mass distribution of $\LbLcPi$ candidates, the normalisation channel in this measurement. The size of the signal peak compared to the backgrounds can be seen clearly in the linear plot on the left, whereas the individual components are displayed more clearly in the logarithmic plot on the right. Overlaid are the fit projections of the signal and background contributions, with individual components illustrated in the legend above.}
	\label{fig:lb2lcpi}
\end{figure}

The $\LcPi$ invariant-mass distribution and the fit projection of the $\LbLcPi$ signal and the background components are shown in Fig.~\ref{fig:lb2lcpi}. The $\LbLcPi$ yield obtained from this fit is $404\,700 \pm 700$, where the uncertainty is statistical. 

\begin{figure}[htp]
   \begin{center}
        \includegraphics[width=0.99\textwidth]{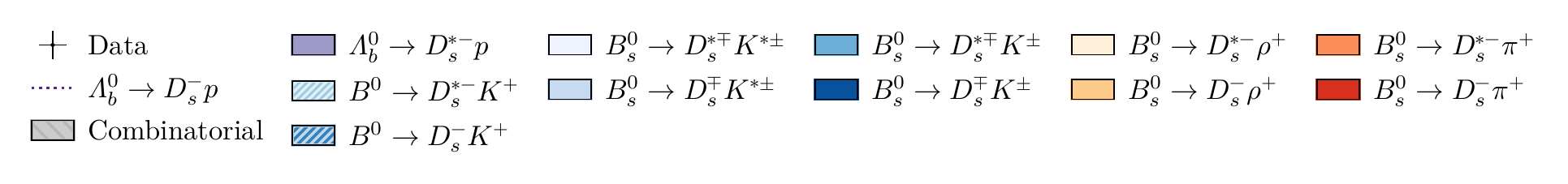}\\
        \includegraphics[width=0.49\textwidth]{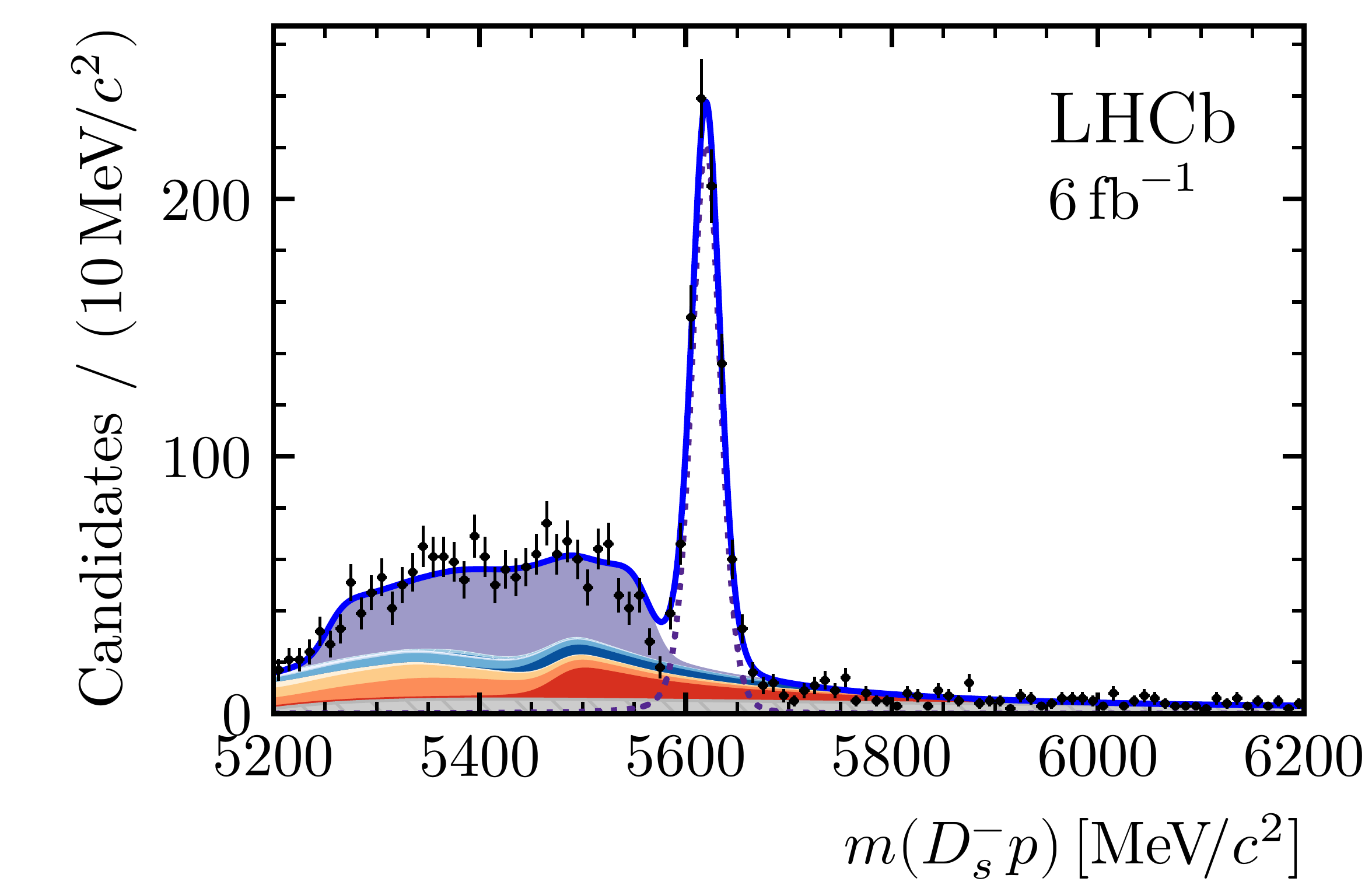}
        \includegraphics[width=0.49\textwidth]{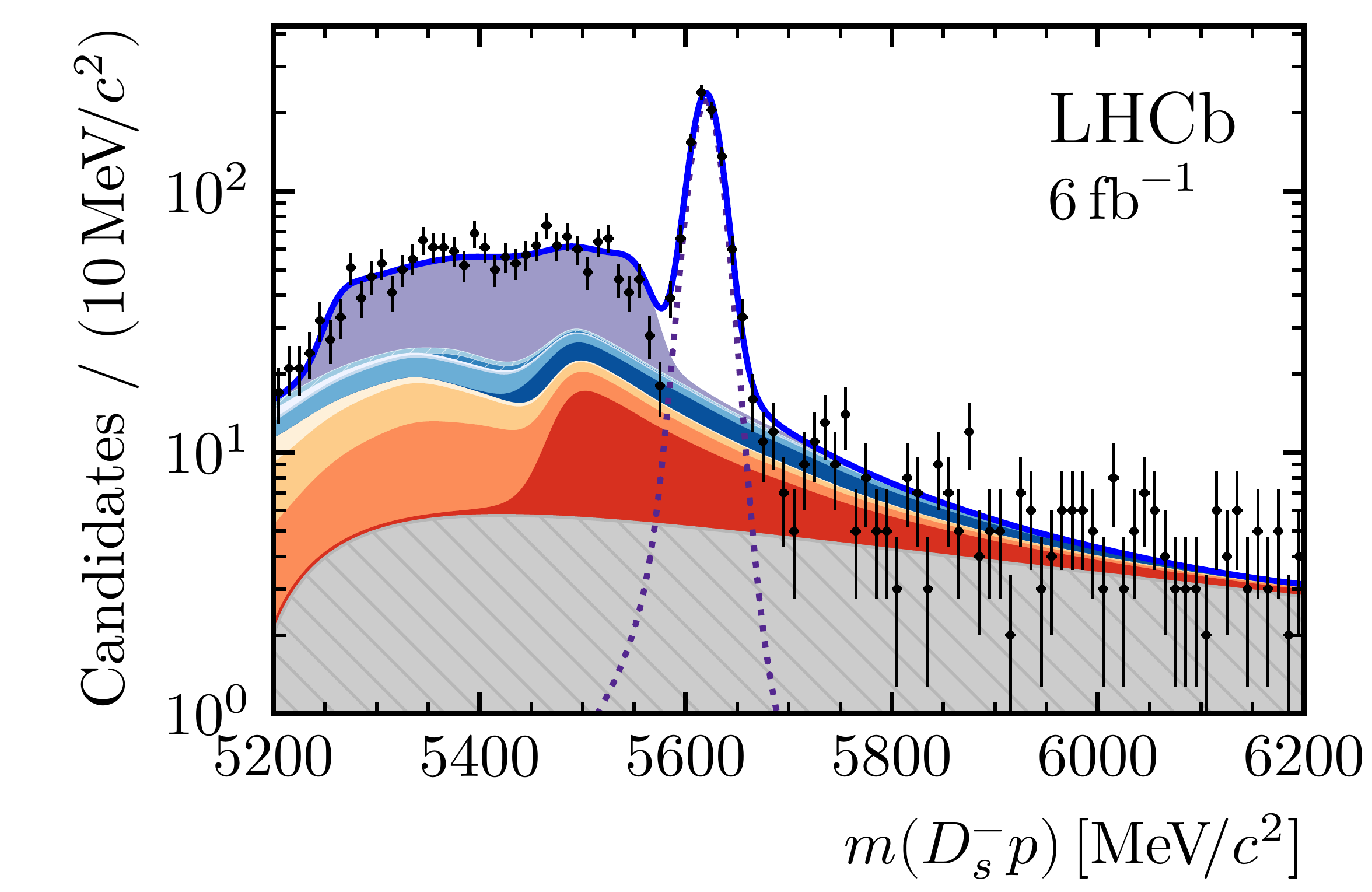}
    \end{center}
	\caption{Invariant-mass distribution of $\LbDsP$ candidates, in (left) linear and (right) logarithmic scale, where the fit projections of the signal and background contributions are overlaid. The individual components in the fit are illustrated in the legend. A few bins around $5575\mevcc$ show a disagreement between the model and the data. This effect is covered by the systematic uncertainties as described in the text.}
	\label{fig:lb2dsp}
\end{figure}

Finally, the fit to the invariant-mass distribution of $\LbDsP$ candidates is shown in Fig.~\ref{fig:lb2dsp}.
A clear $\LbDsP$ signal peak is visible, corresponding to a yield of $\YieldLbDsp$, where the uncertainty is statistical.
This result constitutes the first observation of this decay.
A few bins around $5575\mevcc$ show a disagreement between the model and the data.
This has been scrutinised and any possible effect of this is expected to be covered by systematic uncertainties.
This includes the systematic uncertainties on the shape and yield of the misidentified background components, as discussed in Sec.~\ref{sec:systematics}.
Additionally, the fraction between the two functions used for the shape of the $\LbDsstP$ contribution has been allowed to vary, which has a negligible effect on the signal yield. 

The fits to $\LbDsP$ and $\LbLcPi$ candidates are studied for stability and any bias on the signal yields using pseudoexperiments.
The fits are found to be stable, and no sizeable biases have been found. Furthermore, the fit is validated using data split according to magnet polarity, year of data taking, BDTG response and trigger decision. 

\section{Systematic uncertainties}
\label{sec:systematics}

Systematic uncertainties arising from the limited knowledge of the background and signal shapes, the expected background yields and the PID and hardware trigger efficiencies are considered. 
Due to similarities between the $\LbDsP$ and the $\LbLcPi$ decay topologies, many sources of systematic uncertainties either cancel or are suppressed.
The remaining systematic uncertainties are outlined below and summarised in Table~\ref{tab:syst_uncertainty}.

Systematic uncertainties are assigned to the determination of the $\LbDsP$ and $\LbLcPi$ yields, which are obtained from invariant-mass fits.
These uncertainties are determined using ensembles of pseudoexperiments comparing the results from alternative fit configurations.

For the parametrisation of the signal, the fixed parameters are varied within their uncertainties leading to a variation of $0.54\%$ and $0.27\%$ in the $\LbDsP$ and $\LbLcPi$ yields, respectively.
Systematic uncertainties on the combinatorial background are assessed using pseudoexperiments with alternative parametrisations.
In the fit to $\LbDsP$ candidates, an exponential function is chosen as an alternative to the parametrisation described in the previous section.
The observed variation of $0.73\%$ is assigned as a systematic uncertainty.
A single exponential describes the combinatorial background in the nominal $\mLbLcPi$ fit. In contrast, the sum of two exponential functions is used to assign systematic uncertainties.
This uncertainty amounts to only $0.04\%$ on the final result.

The yields of the misidentified background components in the default invariant-mass fits are constrained or fixed.
The corresponding systematic uncertainty in this assumption is calculated by varying their central values by 10\%, which is twice the uncertainty on the calculated values, to conservatively take into account any bias in the procedure.
This results in a systematic uncertainty of $0.71\%$ and $0.03\%$ in the $\mLbDsP$ and $\mLbLcPi$ fits, respectively.

All the shapes of the misidentified and partially reconstructed background components are varied by applying the PID calibration as a function of different variables.
Samples of $\Dstarp\to\Dz\pip$ and $\Lc\to\proton\Km\pip$ calibration data are used to obtain weighting histograms.
In the default calibration procedure, these histograms depend on the pseudorapidity and momentum of a track.
For systematic studies, they are binned as a function of the number of tracks in the event and the candidate track momentum. 
This variation results in a systematic uncertainty of $0.89\%$ in the sample of $\LbDsP$ candidates and a negligible $0.01\%$ for the $\LbLcPi$ candidates.
Together, the systematic uncertainties on the fit model amount to $1.46\%$ and $0.28\%$ for the fit to $\LbDsP$ and $\LbLcPi$ candidates, respectively.

A systematic uncertainty on the PID selection efficiency is determined by changing the binning scheme of the PID weighting histograms. 
By reducing and increasing the number of bins, the efficiency ratio $\epsilon(\LbLcPi)$/$\epsilon(\LbDsP)$ changes by $0.49\%$, which is therefore assigned as a systematic uncertainty on the final result. This uncertainty mainly originates from the tight PID requirement on the companion proton of the $\LbDsP$ signal.

The efficiency of the hardware trigger requirements is estimated from $\Dstarp$ and $\Lz^0$ calibration samples and using simulation of signal candidates, using a similar method to the one described in Ref.~\cite{LHCb-PAPER-2018-014}.
The difference between data and simulation amounts to $1.15\%$ and is assigned as a systematic uncertainty on the branching fraction result.

Finally, a systematic uncertainty of $0.50\%$ is assigned to the reconstruction efficiency of the charged hadron tracks.
This is mainly due to the uncertainty in simulating the LHCb material correctly and the pion-kaon difference in the final-state particles of the signal and normalisation channels, which have a different interaction cross-section with the material~\cite{LHCb-PAPER-2018-036}.

In summary, the systematic uncertainties in this measurement amount to $2.01\%$ on the final branching fraction result.
This uncertainty is smaller than the statistical uncertainty and the uncertainties arising from the branching fraction inputs.

\begin{table}
    \centering
    \footnotesize
    \caption{Summary of the systematic uncertainties as a percentage of the branching fraction of the $\LbDsP$ decay. The total systematic uncertainty is the quadratic sum of the individual sources.}
    \begin{tabular}{lc}
        \toprule
        Source                          & Relative uncertainty ($\%$) \\
        \midrule
        Invariant-mass fits:            & \\
        \quad $\mLbDsP$ fit:            & \\
        \qquad Signal parametrisation   & 0.54 \\
        \qquad Combinatorial background parametrisation & 0.73 \\
        \qquad Constrained/fixed yields & 0.71 \\
        \qquad Specific background parametrisation & 0.89 \\
        \quad $\mLbLcPi$ fit:           & \\
        \qquad Signal parametrisation   & 0.27 \\
        \qquad Combinatorial background parametrisation & 0.04 \\
        \qquad Constrained/fixed yields & 0.03 \\
        \qquad Specific background parametrisation & 0.01 \\
        Efficiencies:                   & \\
        \quad PID efficiency            & 0.49 \\
        \quad hardware trigger efficiency  & 1.15 \\
        \quad Reconstruction efficiency & 0.50 \\
        \midrule
        Total                           & 2.01 \\
        \bottomrule
    \end{tabular}
    \label{tab:syst_uncertainty}
\end{table}

\section{Results and conclusions}
\label{sec:results}

The branching fraction of $\LbDsP$ can be determined using the efficiencies of the requirements detailed in Sec.~\ref{sec:selection} and the yields of the $\LbDsP$ and $\LbLcPi$ decays as obtained in Sec.~\ref{sec:mass-fits}.
Additionally, inputs for the $\LbLcPi$, $\LcPKPi$ and $\DsKKPi$ branching fractions are required, as defined in Eq.~\eqref{eq:BF}.
Table~\ref{tab:results} shows a summary of the inputs necessary for this measurement.

\begin{table}[b]
    \small
    \centering
    \caption{Obtained signal yields and efficiencies of the $\LbDsP$ and $\LbLcPi$ decays, as well as branching fractions used for this measurement~\cite{PDG2022}. The uncertainty on the signal yields and efficiencies is statistical.}
    \label{tab:results}
    \begin{tabular}{lcc}
    \toprule
                    & $\LbDsP$                & $\LbLcPi$         \\
    \midrule
    Yield           & $\YieldLbDsp $          & $\YieldLbLcPiPaper$    \\
    Efficiency      & $(\EffLbDspPerc)\%$      & $(\EffLbLcPiPerc )\%$   \\
    \midrule       
    $\BF(\LbLcPi)$  & \multicolumn{2}{c}{ $(4.9\phantom{0} \pm 0.4\phantom{0})\times 10^{-3}$ ~\cite{PDG2022}} \\
    $\BF(\DsKKPi)$  & \multicolumn{2}{c}{ $( 5.38 \pm 0.10) \times 10^{-2}$ ~\cite{PDG2022}} \\
    $\BF(\LcPKPi)$  & \multicolumn{2}{c}{ $( 6.28 \pm 0.32 ) \times 10^{-2}$ ~\cite{PDG2022}} \\
    \bottomrule
    \end{tabular}
\end{table}  
The branching-fraction ratio of the $\LbDsP$ and $\LbLcPi$ decays is found to be
\begin{equation*}
    \frac{\BF(\LbDsP)}{\BF(\LbLcPi)} = \RatioBFPAPER \ ,
    \label{BF_ratio}
\end{equation*}
where the first uncertainty is statistical, the second systematic and the third due to the uncertainty of the $\DsKKPi$ and $\LcPKPi$ branching fractions.  

The obtained $\LbDsP$ branching fraction is
\begin{equation*}
    \BF(\LbDsP) = \BFLbDspPaper \ ,
    \label{eq:BF_Lb2Dsp}
\end{equation*}
where the third uncertainty is due to the uncertainty of the $\LbLcPi$, $\DsKKPi$ and $\LcPKPi$ branching fractions.
This measurement is limited by the uncertainty on the $\LbLcPi$ branching fraction, which is dominated by the precision on the ratio of hadronisation fractions $f_{\Lb}/f_{d}$.

In summary, the first observation of the $\LbDsP$ decay and its branching fraction measurement are reported.
Additionally, the branching fraction ratio of the $\LbDsP$ and $\LbLcPi$ decays is determined.
This measurement will serve as input for future studies of factorisation in hadronic $\Lb$ decays.



\clearpage

\section*{Appendices}

\appendix

\section{Particle identification selection on the companion particle}
\label{app:pid-companion}

\begin{figure}[htp]
    \centering
	\includegraphics[width=.475\textwidth]{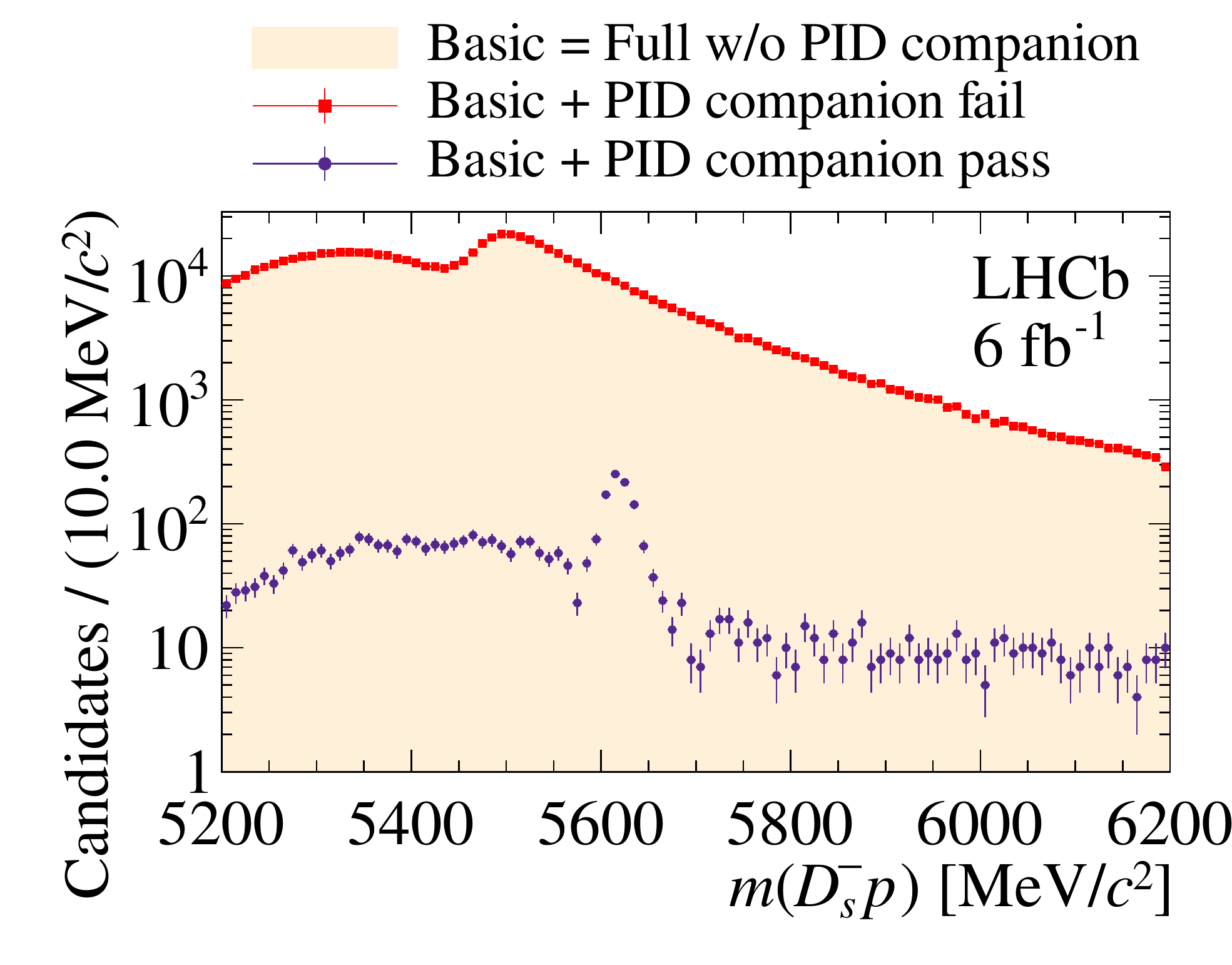}
    \includegraphics[width=.475\textwidth]{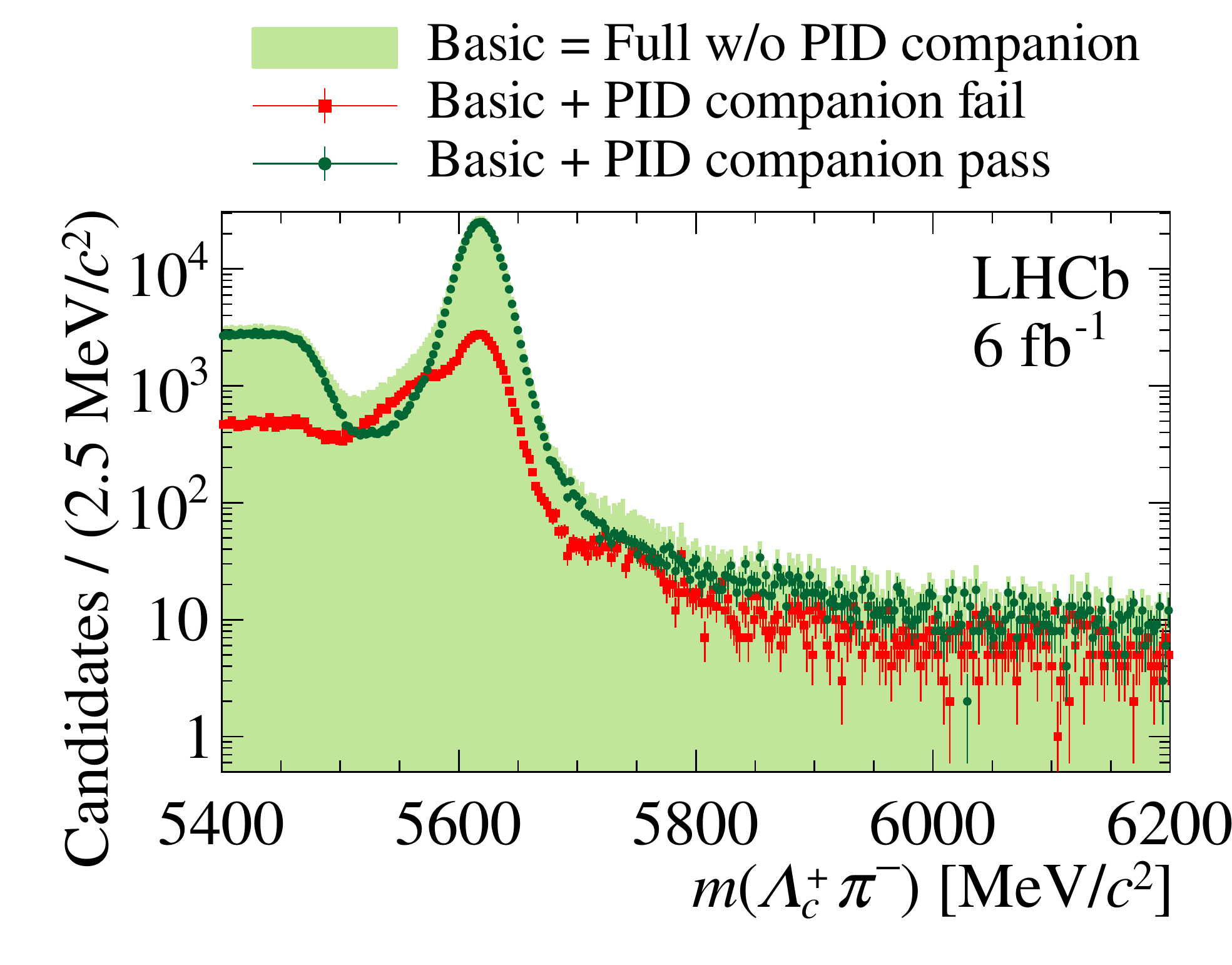} 
	\caption{ The (left) $m(\Dsm p)$ and (right) $\mLbLcPi$ invariant-mass distributions 
	after the full selection without the PID requirement 
	on the companion track (filled area) and passing (circles)
	or failing (squares) this selection.        
		}
    \label{fig:pid-bach}
\end{figure}

Figure~\ref{fig:pid-bach} shows invariant-mass distributions of $\LbDsP$ and $\LbLcPi$ candidates. The candidates are selected by applying the full offline selection without the particle identification (PID) requirement on the companion track. This is represented as the filled orange ($\LbDsP$) and green ($\LbLcPi$) histograms. Candidates additionally passing or failing the PID requirements on the companion track are indicated by circles and squares, respectively. 

\section{Invariant-mass fits to the control channels}
\label{app:control-channels}

\begin{figure}[htp]
   \begin{center}
        \includegraphics[width=0.825\textwidth]{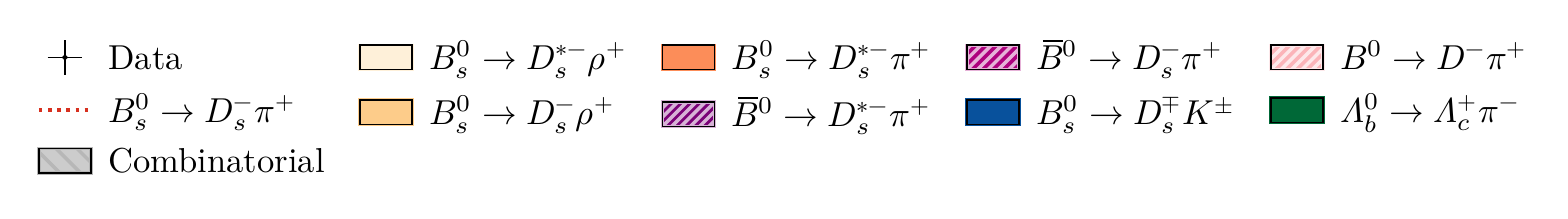}\\
        \includegraphics[width=0.49\textwidth]{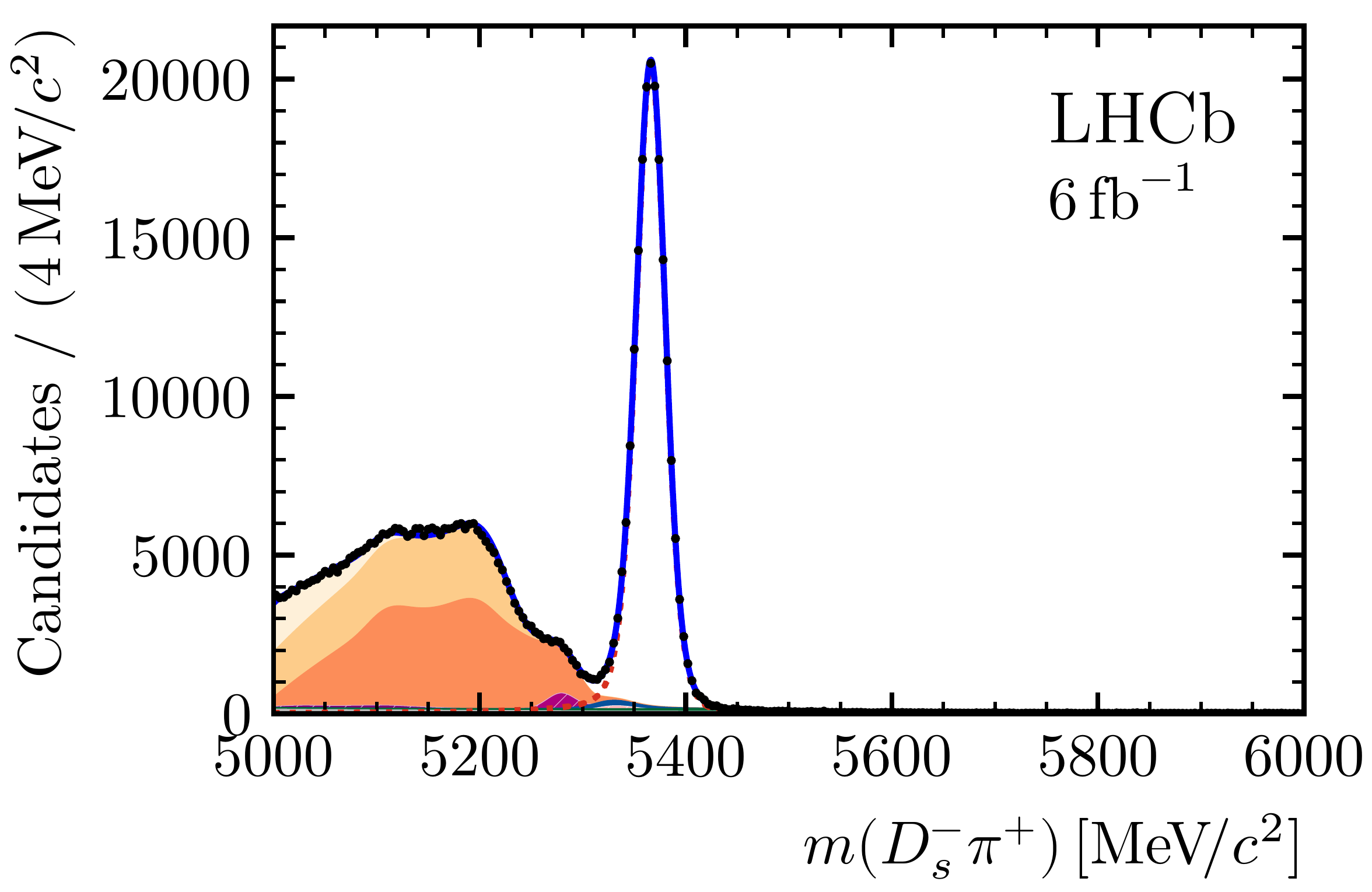}
        \includegraphics[width=0.49\textwidth]{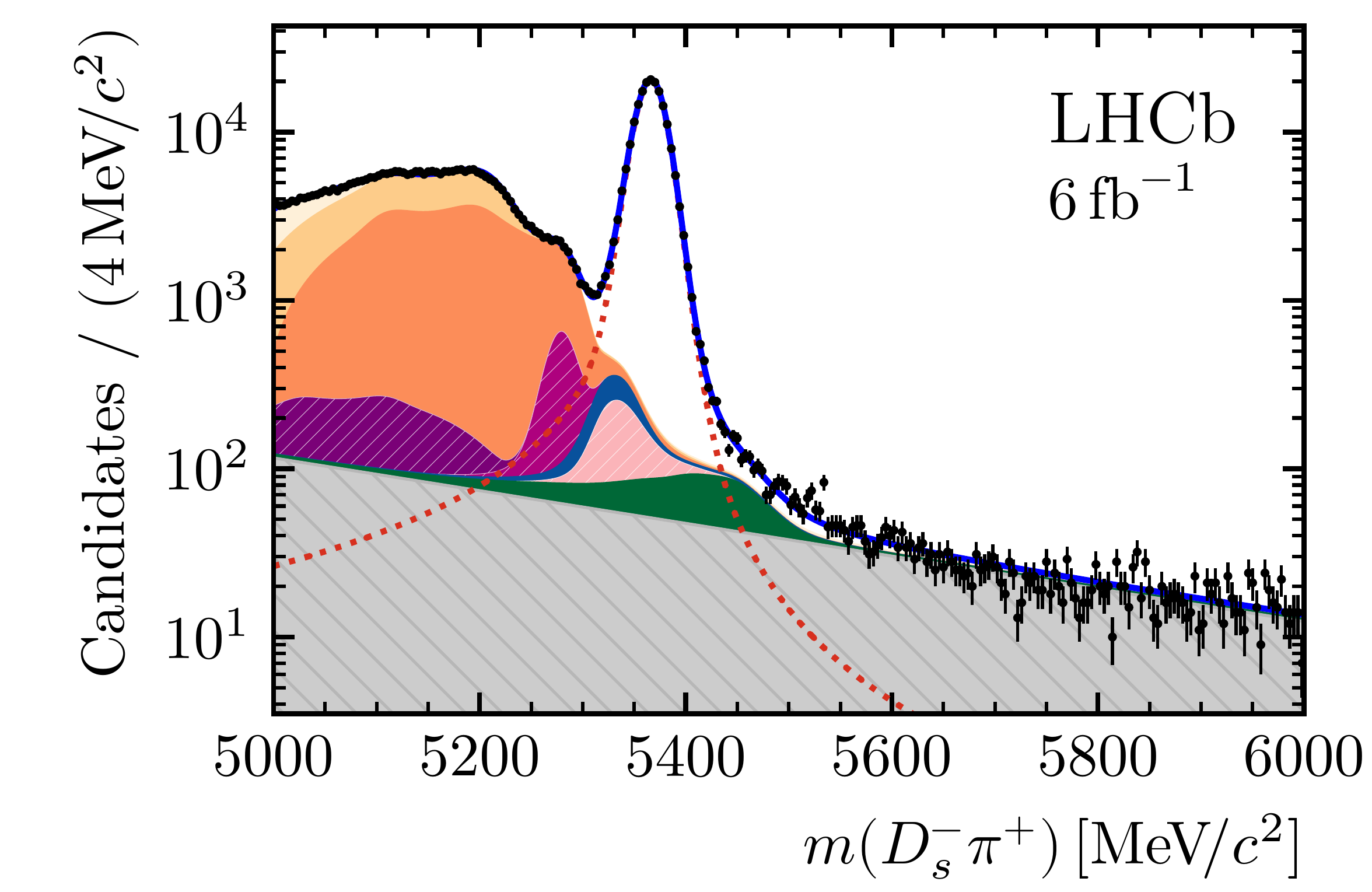}
    \end{center}
	\caption{Invariant-mass distribution of $\BsDsPi$ candidates, in (left) linear and (right) logarithmic scale, where the signal and background contributions are overlaid. The individual components of the fit are illustrated in the legend.}
	\label{fig:bs2dspi}
\end{figure}

To estimate the contribution from $D_s^-\pi^+$-like decays in the \BsDsK and \LbDsP data samples, an invariant-mass fit to the \BsDsPi, \DsKKPi sample under its own hypothesis is performed. The invariant-mass distribution of the \BsDsPi candidates is shown in Fig.~\ref{fig:bs2dspi}. 
The fit model is composed of a signal contribution described by the sum of a double-sided Hypatia function and a Johnson $S_U$ function, combinatorial background described by a single exponential and specific background components. Three different sources of specific background can be distinguished: 
the decays with the same final state as the signal ($\Bz\to\Dsm\pip$), 
the misidentified backgrounds (\BdDPi, \LbLcPi, \BsDsK) and 
the partially-reconstructed backgrounds which occur due to missing particles in the final state (\BdDsstPi, \BsDsstPi, \BsDsRho, \BsDsstRho). The shapes of specific backgrounds are taken from the simulated samples.
The expected yields of \BsDsK, \BdDPi and \LbLcPi decays are calculated from the branching fractions and efficiency predictions.
They are found to be below 1\% of the signal component and are fixed in the fit.
In the default data fit, the signal mean and width, the combinatorial background slope and the yields are left free.

\begin{figure}[htp]
   \begin{center}
        \includegraphics[width=0.99\textwidth]{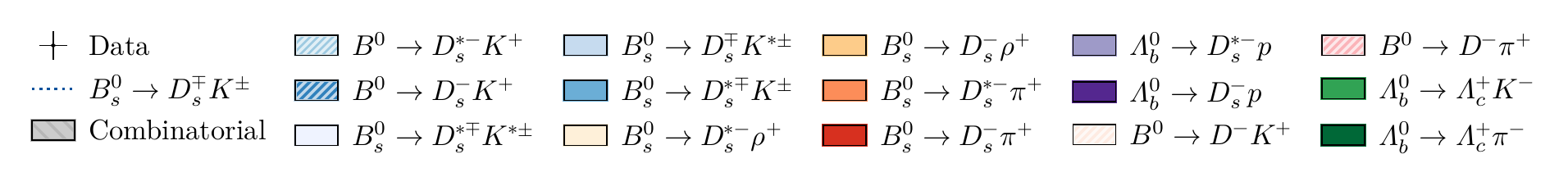}\\
        \includegraphics[width=0.49\textwidth]{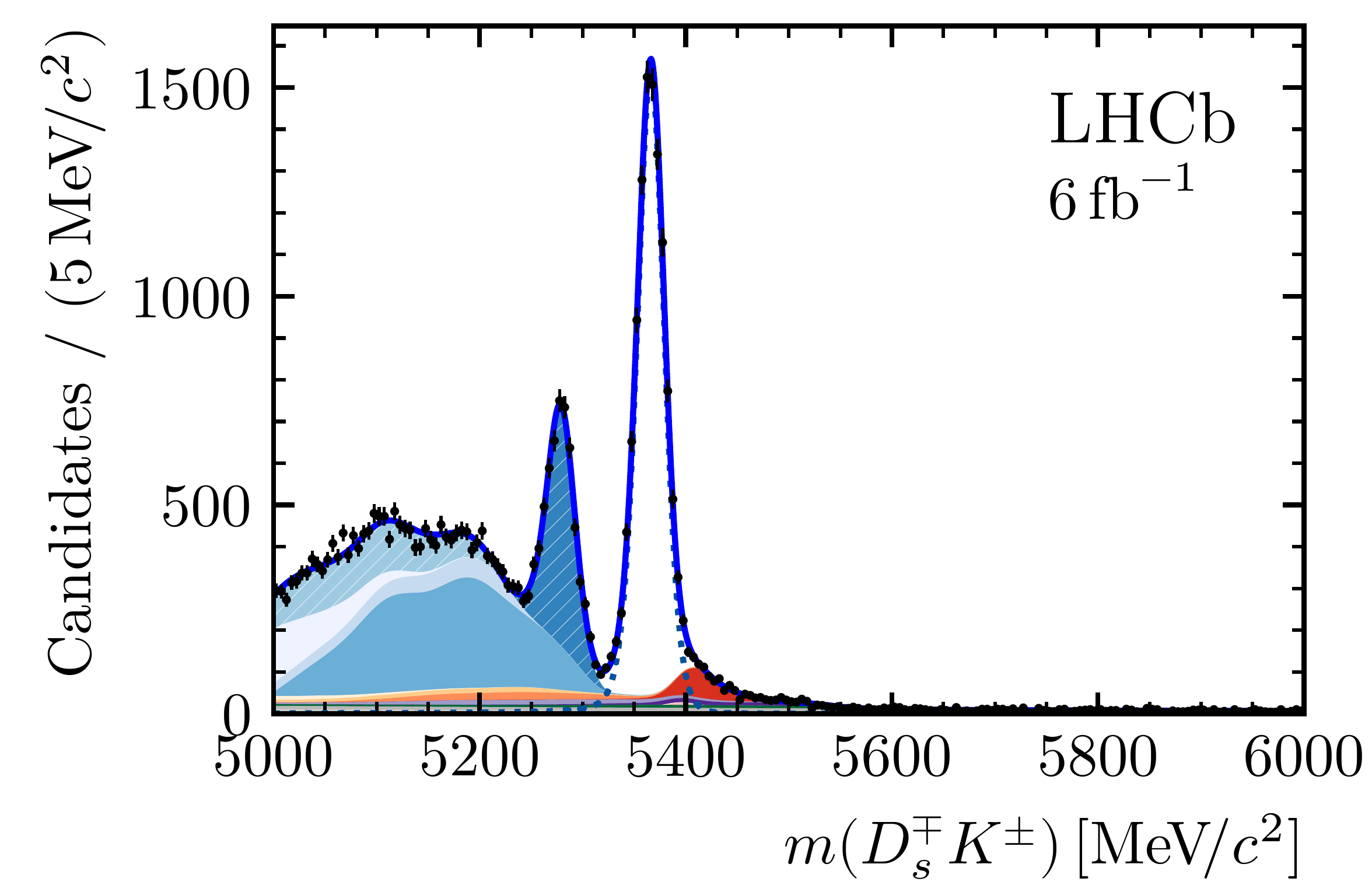}
        \includegraphics[width=0.49\textwidth]{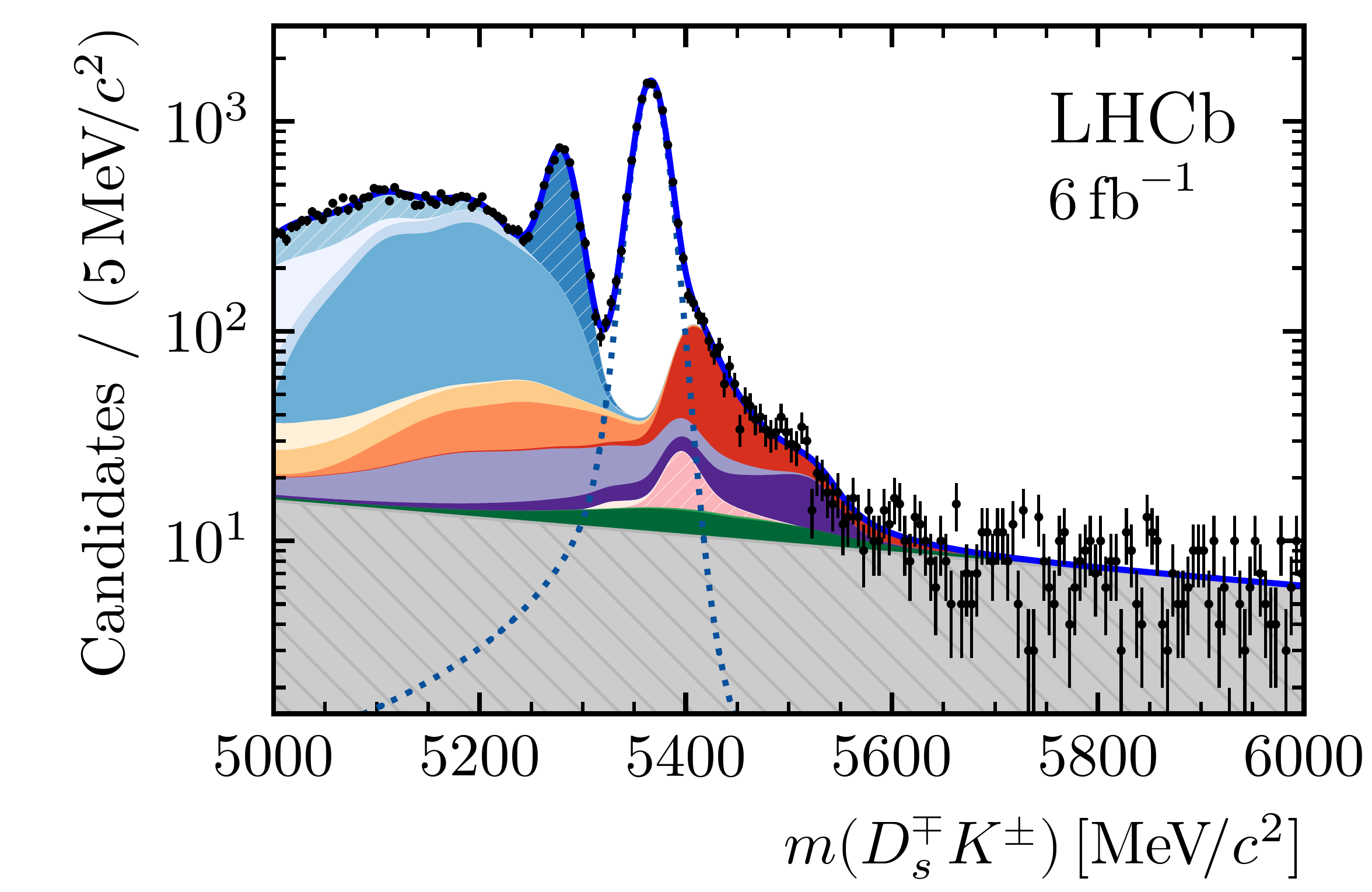}
    \end{center}
	\caption{Invariant-mass distribution of $\BsDsK$ candidates, in (left) linear and (right) logarithmic scale, where the signal and background contributions are overlaid. The individual components in the fit are illustrated in the legend.}
	\label{fig:bs2dsk}
\end{figure}

The contribution from the $D_s^\mp  K^\pm$-like decays in the \LbDsP data sample is estimated from an invariant-mass fit to the \BsDsK sample under its own hypothesis. The invariant-mass distribution of the \BsDsK candidates is shown in Fig.~\ref{fig:bs2dsk}. The fit model is composed of a signal contribution described by the sum of a double-sided Hypatia function and a Johnson $S_U$ function, combinatorial background described by a single exponential and specific background components.
Unlike in the \BsDsPi sample, 
four different sources of background for \BsDsK signal can be distinguished: 
the decays with the same final state as the signal ($\Bz\to\Dsm\Kp$), 
the partially-reconstructed backgrounds which occur due to missing particles in the final state (\BdDsstK, \BsDsstK, \BsDsKst, \BsDsstKst), 
the misidentified backgrounds (\BsDsPi, \LbDsP) and 
the backgrounds that are both partially reconstructed and misidentified (\BsDsstPi, \BsDsRho, \BsDsstRho, \LbDsstP). 
To improve the stability of the fit, 
the contributions from the suppressed \BsDsPilike decays are fixed using the fit to \BsDsPi candidates,
while the remaining yields of the \BdDPi, \BdDK, \LbLcPi and \LbLcK decays are calculated from their branching fractions
and relative efficiencies.
In the default data fit, the signal mean and width, the combinatorial background slope and the yields are left free. 

\clearpage


\section*{Acknowledgements}
%
%
\noindent We express our gratitude to our colleagues in the CERN
accelerator departments for the excellent performance of the LHC. We
thank the technical and administrative staff at the LHCb
institutes.
We acknowledge support from CERN and from the national agencies:
CAPES, CNPq, FAPERJ and FINEP (Brazil); 
MOST and NSFC (China); 
CNRS/IN2P3 (France); 
BMBF, DFG and MPG (Germany); 
INFN (Italy); 
NWO (Netherlands); 
MNiSW and NCN (Poland); 
MEN/IFA (Romania); 
MICINN (Spain); 
SNSF and SER (Switzerland); 
NASU (Ukraine); 
STFC (United Kingdom); 
DOE NP and NSF (USA).
We acknowledge the computing resources that are provided by CERN, IN2P3
(France), KIT and DESY (Germany), INFN (Italy), SURF (Netherlands),
PIC (Spain), GridPP (United Kingdom), 
CSCS (Switzerland), IFIN-HH (Romania), CBPF (Brazil),
Polish WLCG  (Poland) and NERSC (USA).
We are indebted to the communities behind the multiple open-source
software packages on which we depend.
Individual groups or members have received support from
ARC and ARDC (Australia);
Minciencias (Colombia);
AvH Foundation (Germany);
EPLANET, Marie Sk\l{}odowska-Curie Actions and ERC (European Union);
A*MIDEX, ANR, IPhU and Labex P2IO, and R\'{e}gion Auvergne-Rh\^{o}ne-Alpes (France);
Key Research Program of Frontier Sciences of CAS, CAS PIFI, CAS CCEPP, 
Fundamental Research Funds for the Central Universities, 
and Sci. \& Tech. Program of Guangzhou (China);
GVA, XuntaGal, GENCAT and Prog.~Atracci\'on Talento, CM (Spain);
SRC (Sweden);
the Leverhulme Trust, the Royal Society
 and UKRI (United Kingdom).

\addcontentsline{toc}{section}{References}
\bibliographystyle{LHCb}
\bibliography{main,standard,LHCb-PAPER,LHCb-CONF,LHCb-DP,LHCb-TDR}

\newpage
\centerline
{\large\bf LHCb collaboration}
\begin
{flushleft}
\small
R.~Aaij$^{32}$\lhcborcid{0000-0003-0533-1952},
A.S.W.~Abdelmotteleb$^{50}$\lhcborcid{0000-0001-7905-0542},
C.~Abellan~Beteta$^{44}$,
F.~Abudin{\'e}n$^{50}$\lhcborcid{0000-0002-6737-3528},
T.~Ackernley$^{54}$\lhcborcid{0000-0002-5951-3498},
B.~Adeva$^{40}$\lhcborcid{0000-0001-9756-3712},
M.~Adinolfi$^{48}$\lhcborcid{0000-0002-1326-1264},
P.~Adlarson$^{77}$\lhcborcid{0000-0001-6280-3851},
H.~Afsharnia$^{9}$,
C.~Agapopoulou$^{13}$\lhcborcid{0000-0002-2368-0147},
C.A.~Aidala$^{78}$\lhcborcid{0000-0001-9540-4988}
% ...

{\footnotesize \it

% ...
$^{9}$Universit{\'e} Clermont Auvergne, CNRS/IN2P3, LPC, Clermont-Ferrand, France\\
% ...
$^{13}$LPNHE, Sorbonne Universit{\'e}, Paris Diderot Sorbonne Paris Cit{\'e}, CNRS/IN2P3, Paris, France\\
% ...
$^{32}$Nikhef National Institute for Subatomic Physics, Amsterdam, Netherlands\\
% ...
$^{40}$Instituto Galego de F{\'\i}sica de Altas Enerx{\'\i}as (IGFAE), Universidade de Santiago de Compostela, Santiago de Compostela, Spain\\
% ...
$^{44}$Physik-Institut, Universit{\"a}t Z{\"u}rich, Z{\"u}rich, Switzerland\\
% ...
$^{48}$H.H. Wills Physics Laboratory, University of Bristol, Bristol, United Kingdom\\
% ...
$^{50}$Department of Physics, University of Warwick, Coventry, United Kingdom\\
% ...
$^{53}$School of Physics and Astronomy, University of Glasgow, Glasgow, United Kingdom\\
$^{54}$Oliver Lodge Laboratory, University of Liverpool, Liverpool, United Kingdom\\
% ...
$^{62}$Syracuse University, Syracuse, NY, United States\\
% ...
$^{77}$Department of Physics and Astronomy, Uppsala University, Uppsala, Sweden, associated to $^{53}$\\
$^{78}$University of Michigan, Ann Arbor, MI, United States, associated to $^{62}$\\
\bigskip
% ...
}
\end{flushleft}




\end{document}